\documentclass[aps,prb,showpacs,twocolumn,amsmath,amssymb,superscriptaddress]{revtex4}

\usepackage{graphicx}
\usepackage{multirow}
\usepackage{dcolumn}
\usepackage{bm}
\usepackage{float}

\begin{document}

\title{Gate-Tunable Exchange Coupling Between Cobalt Clusters on Graphene}

\author{Hua Chen}
    \affiliation{Department of Physics, University of Texas at Austin, Austin, TX 78712, USA}

\author{Qian Niu}
    \affiliation{Department of Physics, University of Texas at Austin, Austin, TX 78712, USA}

\author{Zhenyu Zhang}
   \affiliation{ICQD/HFNL, University of Science and Technology of China, Hefei, Anhui, 230026, China}
\date{\today}

\author{Allan H. MacDonald}
    \affiliation{Department of Physics, University of Texas at Austin, Austin, TX 78712, USA}
    
\begin{abstract}
We use spin-density-functional theory (SDFT) {\em ab initio} calculations to theoretically explore the possibility of achieving useful gate control over exchange coupling between cobalt clusters placed on a graphene sheet. By applying an electric field across supercells we demonstrate that the exchange interaction is strongly dependent on gate voltage, but
find that it is also sensitive to the relative sublattice registration of the cobalt clusters. We use our results to discuss strategies for achieving strong and reproducible magneto-electric effects in graphene/transition-metal hybrid systems.    
\end{abstract}


\maketitle

\section{Introduction}
Graphene\cite{geim_2007,castroneto_2009} is an atomically thin two-dimensional gapless semi-conductor in which the carrier density can be varied over a broad range, from $\sim - 10^{13}$ cm$^{-2}$ to $\sim + 10^{13}$ cm$^{-2}$ by gating, and is a remarkably good conductor at high carrier densities. Graphene/transition metal hybrid systems are attractive for spintronics because carbon spin-orbit interactions are particularly weak\cite{min_2006,pesin_2012} in flat honeycomb-lattice arrays, because magnetic transition element clusters\cite{baumer_1995,weser_2010,vovan_2010} form readily on graphene surfaces, and because of potentially attractive properties \cite{karpan_2007,karpan_2008} of interfaces between graphene and magnetic transition metals. For example ultra-thin transition metal layers on graphene are predicted\cite{gong_2012,porter_2012} to have extremely large magnetic metal anisotropy energies. For these reasons there has recently been considerable interest\cite{xiao_2009,johll_2009,wehling_2011,cheianov_2006,saha_2012} in the magnetic and electronic properties of transition metal adatoms and clusters placed on a two-dimensional graphene sheet. 

In this article we theoretically explore the possibility that the exchange coupling between separate magnetic metal clusters on graphene can be altered electrically by gating. Since arrays of magnetic clusters can be realized on graphene by using a graphene/substrate moir{\' e} pattern\cite{vovan_2011} as a template, and the magnetic clusters hybridize relatively strongly with graphene's valence and conduction band orbitals, we anticipate gate-dependent exchange coupling between clusters which should lead to gate-dependent magneto-resistance\cite{baibich_1988,binasch_1989} effects that are strong at room temperature. The goal of this work is to identify strategies for achieving strong, reproducible magneto-electric effects in graphene/transition-metal hybrid systems.  

There is already a substantial theoretical literature\cite{vozmediano_2005, dugaev_2006, saremi_2007, brey_2007, bunder_2009, black-schaffer_2010, sherafati_2011_1, sherafati_2011_2, kogan_2011} on Ruderman-Kittel-Kasuya-Yosida (RKKY) interactions between local moments coupled to graphene $\pi$-bands. It has been recognized, \cite{saremi_2007} for example, that when graphene is undoped the RKKY interaction is ferromagnetic (FM) for magnetic moments coupled to $\pi$-electrons on the same graphene sublattice and antiferromagnetic (AFM) for moments coupled to $\pi$-electrons on different sublattices. The RKKY interaction decays as $r^{-3}$ at large distance $r$, because of the suppressed density-of-states at the Dirac point of graphene \cite{saremi_2007, brey_2007, sherafati_2011_1, kogan_2011}. At finite carrier density the RKKY coupling has spatial oscillations with period $\pi/k_F$ on top of an envelope which decays as $r^{-2}$. Most existing studies of the RKKY interactions in graphene have assumed magnetic moments due to point-like impurities that are associated with a particular honeycomb lattice site and have purely phenomenological interactions. These models are realized approximately in systems with magnetic moments due to hydrogenation \cite{zhou_2009} or carbon vacancies \cite{chen_2011}, although these defects significantly modify the carbon $sp^2$ bonds and hence the structural and electronic properties of graphene. Moments due to adsorbed magnetic transition metal atoms do not 
distort the graphene bands as strongly but have small migration barriers \cite{ding_2011} due to weak adsorption energies. \cite{johll_2009}  The transition metal clusters on graphene on which we focus are relatively immobile, 
however, and can be large enough to exceed the super paramagnetic limit. These larger magnetic objects therefore have more potential for spintronics applications. We attempt to realistically describe the magnitude of cobalt cluster moments, their magnetic anisotropy energies (MAE), the exchange coupling between
the clusters and graphene, and finally the graphene-mediated magnetic exchange energies between separated clusters.  

We use first-principles supercell electronic structure calculations based on spin density functional theory (SDFT) to investigate not only the RKKY coupling between magnetic cobalt clusters deposited on graphene, but also its dependence on external electric fields due to gating. We choose cobalt  because its bulk lattice constant is very close to that of graphene, and because thin cobalt films down to two or three atomic layers have been found to have perpendicular magnetic anisotropy \cite{vovan_2010}, which is preferable for spintronic applications. First, by calculating the electronic structure of a two-atomic-layer thick cobalt film on graphene, we find that there is considerable charge transfer from cobalt to graphene. Hybridization between the cobalt cluster and graphene leads to sublattice and spin dependent shifts in graphene $\pi$-band energies from which we are able to extract the essential kinetic-exchange parameters. Then we directly calculate the exchange interaction between two parallel two-atomic-layer-thick cobalt  ribbons placed on graphene. For the geometries we have been able to consider, we find that the exchange interactions have a typical size $\sim 10^{-4}$ eV per cobalt atom, comparable to the MAE of bulk cobalt ($4\times 10^{-5}$ eV \cite{bate_1991}) and thin films of cobalt on graphene \cite{vovan_2010}, but smaller than anisotropy energies which can be achieved in asymmetrical clusters.\cite{gambardella_2002, canali_2007} We also find that exchange interaction tend to change sign when a graphene cluster changes its sublattice registration, and that the exchange interactions can be modified by gate voltages.

In Section II we briefly describe the methods that we use for these computations. For the sake of definiteness we have focused our attention on cobalt clusters that are two atomic layers thick and arranged in a ribbon geometry. In Section III we describe our results for the electronic structure of a bulk two-layer thick film of cobalt on graphene. We find that there is considerable charge transfer from cobalt to graphene, and that hybridization between the magnetic cluster and graphene leads to sublattice and spin dependent shifts in graphene $\pi$-band energies. In Section IV we summarize our results for the dependence of total energy on the relative spin orientations of separated clusters. We are able to understand our main findings using an approximate treatment which treats the cobalt-graphene interaction
perturbatively. Finally in Section V we present our results for the gate-voltage dependence of these exchange interactions. We find that gate fields can produce sizable changes in exchange interactions, in some cases changing their signs and substantially reducing their sublattice registration dependence. In Section VI we summarize our findings and discuss some possible directions for future research.

\section{Methods} 

The DFT calculations reported on in this work were performed using the projector-augmented-wave (PAW) \cite{blochl_1994} method as implemented in the Vienna {\it ab initio} simulation package (VASP) \cite{kresse_1996_1,kresse_1996_2,kresse_1999}. The Perdew-Burke-Emzerhof generalized gradient approximation (PBE-GGA) \cite{perdew_1996} was used for the exchange-correlation energy functional. To calculate the electronic band structure of an infinite graphene sheet fully covered by a two-atomic-layer-thick cobalt  film [Fig.~\ref{figure1} (a)], we used a  a 20 \r{A} thick vacuum region between neighboring supercells in the $\hat{z}$ (perpendicular to the graphene plane) direction. 
We fixed the lattice constant at the experimental value for graphene (2.46 \r{A})since the (0001) surface of bulk hcp cobalt has a small lattice mismatch ($<2\%$). All atoms in the supercell were allowed to relax until the Hellmann-Feynman force on each atom was smaller than 0.001 eV/\r{A}. A plane-wave energy cutoff of 400 eV and a $33\times 33\times 1$ $k$-point mesh were used for structure relaxation and total energy calculations. Denser $k$-point meshes (up to $79\times 79\times 1$) were used to check accuracy and to perform MAE calculations.

To study the indirect exchange coupling between remote cobalt clusters on graphene, we constructed a supercell with two parallel cobalt ribbons two-atomic-layers thick and three atoms wide, oriented along the zigzag direction of graphene (Fig.~\ref{figure2}). The supercell used in this case is $25\times 1$ with the same 20 \r{A} vacuum layer in $\hat{z}$ direction. These ribbon calculations used a $1\times 49\times 1$ $k$-point mesh. The lattice parameters of the cobalt ribbons were taken from the infinite 2D slab calculations mentioned above without further relaxation. (We checked the influence of relaxation for several cases and did not find qualitative modification relative to the results reported on below.) The exchange coupling between the cobalt ribbons was estimated by calculating the total energy difference between spin-parallel (FM) and spin-antiparallel (AFM) configurations:
\begin{equation}\label{eq:defcoup}
\Delta E=E_{\text{FM}}-E_{\text{AFM}}.
\end{equation}
With this convention a positive $\Delta E$ corresponds to antiferromagnetic exchange between the ribbons.

An external electric field across the supercells was realized by adding a saw-tooth like external potential to the total energy functional \cite{neugebauer_1992}. We have applied electric fields of different size in the same supercell as in Fig.~\ref{figure2}. In this case the external field can produce only charge transfer between the two cobalt ribbons and graphene. A more realistic representation of gating action on the graphene/transition metal hybrid system can be achieved by adding a bilayer Cu slab to the supercell as in (Fig.~\ref{figure9}). The copper acts as a a charge reservoir and also screens the part of graphene directly below the cobalt ribbons from external fields. A more detailed discussion of some issues involved in using VASP to simulate gates is provided in Appendix B.  

\section{Kinetic Exchange coupling between cobalt overlayers and graphene $\pi$-bands}

\subsection{{\em Ab Initio} Spin-density-functional Theory}

\begin{figure}[h]
 \begin{center}
\includegraphics[width=3.4 in]{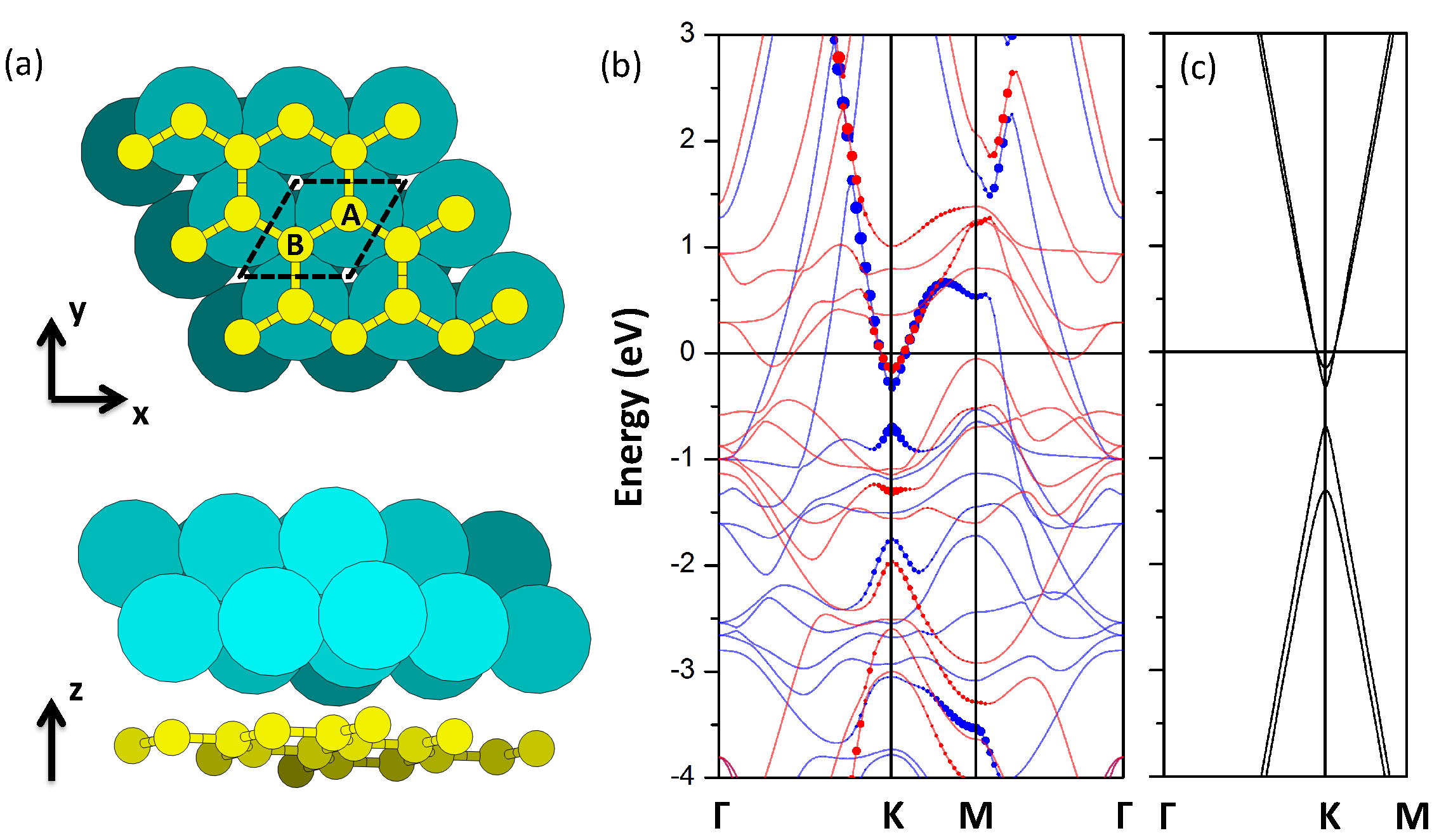}
 \end{center}
 \caption{(color online). (a) Top and side views of the supercell (with a $3\times 3$ repetition in the $xy$ plane for illustration purpose). The larger balls represent cobalt atoms and the smaller balls C atoms. (b) Two-dimensional Kohn-Sham quasiparticle band structure of the Co-graphene hybrid system neglecting spin-orbit interactions. The blue lines illustrate the majority spin bands and the red lines the minority spin bands. The blue and red dots indicate the strength of carbon $p_z$ orbital character in the majority and minority spin states. (c) Model graphene projected band structure calculated using Eq.~\ref{eq:model}. The model parameter values (Eq.~\ref{eqn:modelpara}) are obtained by fitting to the DFT results listed in Table~\ref{table1}.}
\label{figure1}
\end{figure}

As illustrated in Fig.~\ref{figure1} (a), we have calculated the total energies of bilayer cobalt films adsorbed on graphene with different registries and have found that the most stable geometry is that with the C atoms in one sublattice of graphene located directly below bottom-layer cobalt atoms, {\em i.e.} at atop sites, and the C atoms in the other sublattice below the top-layer cobalt atoms, {\em i.e.} at hcp sites. The optimal separation 
between the cobalt overlayer and graphene is about 2.21 \r{A}. After adsorbtion on graphene, the magnetic moments on the cobalt atoms in the first layer (adjacent to graphene) decrease from 1.710 $\mu_{\text{B}}$ per cobalt atom, which is close to the bulk value, to 1.560 $\mu_{\text{B}}$ per cobalt atom. Meanwhile, the C atoms in sublattice A (adjacent to cobalt atoms) obtain a per-atom magnetic moment of 0.043 $\mu_{\text{B}}$, antiparallel to the magnetization of the cobalt overlayer, whereas the C atoms in sublattice B acquire a moment of 0.041 $\mu_{\text{B}}$ per atom and parallel to the cobalt moments. Therefore the overall magnetization direction of graphene is opposite to that of the cobalt film. We have also calculated the magnetocrystalline part of the MAE by evaluating the total energy difference, including spin-orbit coupling, between configurations with all moments along the $\hat{z}$ direction (out-of-plane) and along the $\hat{x}$ direction (in-plane). The system is found to have perpendicular magnetic anisotropy \cite{vovan_2010}, with a MAE of $\sim$0.09 meV per cobalt atom, which is larger than that of bulk hcp cobalt ($\sim$0.04 meV), but still the same order of magnitude.

The spin-resolved Kohn-Sham band structure of the Co-graphene hybrid system is shown in Fig.~\ref{figure1} (b). The graphene bands are spin-split and the Dirac points at the $K$ point are gapped because of the relatively strong interaction with the cobalt overlayer, in agreement with previous results \cite{giovannetti_2008,khomyakov_2009,eom_2009}. It is nevertheless clear from the position of the Fermi level that graphene is $n$-doped, {\em i.e.} electrons are transferred from cobalt to graphene \cite{khomyakov_2009}. The graphene layer majority-spin Dirac point is easily identified in the two-dimensional bands, but its minority spin-counterpart is so strongly hybridized with cobalt $d$-orbitals that it is less easily identified. The $K$ point is at a higher energy for graphene majority spin bands than for minority spin bands, indicating an overall antiferromagnetic coupling between the cobalt overlayer and graphene. This conclusion is also in agreement with the antiparallel orientations of the graphene and cobalt magnetizations mentioned above. 

\subsection{Kinetic Exchange Model}  

Our electronic structure calculations can be qualitatively described using a simple model for graphene coupled to a cobalt overlayer in which hybridization and charge transfer effects shift the energies of both majority and minority spins on both graphene sublattices:
\begin{eqnarray}
H=\hbar v_F \bm k \cdot \bm \tau + \mu - h_{0,z} \tau_z - h_{z,0} S_{z} - h_{z,z} S_{z} \tau_z.
\label{eq:model} 
\end{eqnarray}
In Eq.~\ref{eq:model} the first term on the right hand side is the usual Dirac Hamiltonian for hopping on a honeycomb lattice with velocity $v_{F} \sim 10^{6}$ m/s and wave vectors measured relative to the Brillouin-zone corners, $S_{z} = \pm 1/2$ labels spin, and $\tau_z = \pm 1$ distinguishes A (under the atop site) and B (under the hcp site) sublattices. The parameters of this model can be identified by fitting to the energies of the bands that have the largest $\pi$-band character at the Brillouin-zone corner ($K$) points, which are summarized in Table ~\ref{table1}. 
$H$ is diagonal when $\bm k=0$ and its four eigenvalues
\begin{eqnarray}
&&\mu - h_{0,z}-\frac{1}{2}h_{z,0}-\frac{1}{2}h_{z,z},\\\nonumber 
&&\mu - h_{0,z}+\frac{1}{2}h_{z,0}+\frac{1}{2}h_{z,z},\\\nonumber
&&\mu + h_{0,z}-\frac{1}{2}h_{z,0}+\frac{1}{2}h_{z,z},\\\nonumber
&&\mu + h_{0,z}+\frac{1}{2}h_{z,0}-\frac{1}{2}h_{z,z},
\end{eqnarray}
correspond to the four eigenvectors
\begin{eqnarray}
|A\uparrow\rangle, |A\downarrow\rangle, |B\uparrow\rangle, |B\downarrow\rangle.
\end{eqnarray}
The four Kohn-Sham bands with the strongest carbon $p_z$ character at the $K$ point of Brillouin
zone are bands 2, 3, 4, and 5 in Table ~\ref{table1}. 
By fitting their energies to the SDFT band energies we can obtain the values of the parameters:
\begin{eqnarray}
\mu &=& -0.622\mbox{ eV}\\\nonumber
h_{0,z} &=& 0.195\mbox{ eV}\\\nonumber
h_{z,0} &=& -0.214\mbox{ eV}\\\nonumber
h_{z,z} &=& -0.766\mbox{ eV}.
\label{eqn:modelpara}
\end{eqnarray}
The model band structure calculated with these these parameters is plotted in Fig.~\ref{figure1} (c).

Several comments are in order:

\noindent 
(i) The chemical potential $\mu$ specifies the energy shift averaged over spin and sublattice, which is negative because electrons are transferred to graphene, in agreement with our previous discussion. 

\noindent
(ii) The value of $h_{0,z}$ is positive because the A sublattice is more strongly influenced by the cobalt overlayer than the B sublattice, which is expected since the A sublattice is directly below the cobalt atoms at the interface.

\noindent
(iii) The value of $h_{z,0}$ measures the kinetic exchange coupling between cobalt and graphene spins averaged over sublattices. Its negative sign means the sublattice-averaged magnetic coupling is AFM, also in agreement with our observations in the previous subsection. 

\noindent
(iv) The spin- and sublattice-dependent term $h_{z,z}$ reflects the property that the majority spin is higher in energy on the A sublattice whereas the minority spin is higher in energy on the B sublattice. In other words, the Co-graphene exchange coupling is AFM on the A sublattice but FM on the B sublattice. To understand this property we refer to Table~\ref{table1} in which bands 1, 6, 7, 8 are identified as cobalt $d$ bands that hybridize with the graphene $\pi$ bands. From the carbon $p_z$ and cobalt $d$ characters that these bands carry, it can be seen that spin-splitting on the A sublattice is because of hybridization mainly with the $d^{3z^2-r^2}$ orbitals of cobalt (bands 1 and 6), whose minority spin states are higher in energy than majority spin states and above the Fermi level. The higher energy of the carbon majority spin states on the A sublattice can therefore be understood as the result of level repulsion from cobalt $d^{3z^2-r^2}$ orbitals with the same spin. The same argument also applies for the B sublattice, whose $p_z$ orbitals mainly hybridize with the $d^{xz}$, $d^{yz}$, $d^{xy}$, and $d^{x^2-y^2}$ orbitals of cobalt because of symmetry. However, both of the two cobalt $d$ bands (band 7 and 8) with these characterare below the Fermi energy and the $\pi$-bands at the $K$ point, with the minority spin band higher in energy. Therefore level repulsion in this case results in the higher energy of the minority spin states, {\em i.e.} in ferromagnetic coupling.  

\noindent
(v) $h_{z,z}$ is much larger than $h_{z,0}$ because the kinetic exchange interaction between the cobalt overlayer and the graphene is strongly dependent on sublattice. We will see later that this property will translate to a strong dependence of the graphene-mediated exchange interaction between two cobalt clusters on their relative registries with respect to the sublattices of a continuous graphene sheet.

\begin{table}[h]
\renewcommand{\arraystretch}{1.5} 
  \caption{Orbital character of the bands in Fig.~\ref{figure1} (b) at the $K$ point of 2D Brillouin zone. Only those having strong carbon $p_z$ characters are listed. A and B correspond to the two sublattices of graphene, as shown in Fig.~\ref{figure1} (a).}
\label{table1}
\begin{tabular}{llll}
\hline\hline
 Band No. & Energy (eV) & C $p_z$ character & cobalt $d$ character \\\hline
 1        & 1.006       & A$\downarrow$: 0.069 & $3z^2-r^2\downarrow$: 0.671\\\hline
 \multirow{2}{*}{2} & \multirow{2}{*}{-0.151} & \multirow{2}{*}{B$\downarrow$: 0.319} & $xz,yz\downarrow$: 0.098 \\
&&&$x^2-y^2,xy\downarrow$: 0.099\\\hline
 3        & -0.328       & A$\uparrow$: 0.297 & $3z^2-r^2\uparrow$: 0.368\\\hline
 \multirow{2}{*}{4} & \multirow{2}{*}{-0.703} & \multirow{2}{*}{B$\uparrow$: 0.439} & $xz,yz\uparrow$: 0.059 \\
&&&$x^2-y^2,xy\uparrow$: 0.016\\\hline
 5        & -1.307       & A$\downarrow$: 0.341 & $3z^2-r^2\downarrow$: 0.019\\\hline
 6        & -1.754       & A$\uparrow$: 0.191 & $3z^2-r^2\uparrow$: 0.303\\\hline
 \multirow{2}{*}{7} & \multirow{2}{*}{-1.965} & \multirow{2}{*}{B$\downarrow$: 0.185} & $xz,yz\downarrow$: 0.19 \\
&&&$x^2-y^2,xy\downarrow$: 0.054\\\hline
 \multirow{2}{*}{8} & \multirow{2}{*}{-3.048} & \multirow{2}{*}{B$\uparrow$: 0.055} & $xz,yz\uparrow$: 0.206 \\
&&&$x^2-y^2,xy\uparrow$: 0.166\\
\hline\hline
\end{tabular}
\end{table}

\section{Magnetic coupling between cobalt clusters on neutral graphene}

\begin{figure}[h]
 \begin{center}
\includegraphics[width=3.4 in]{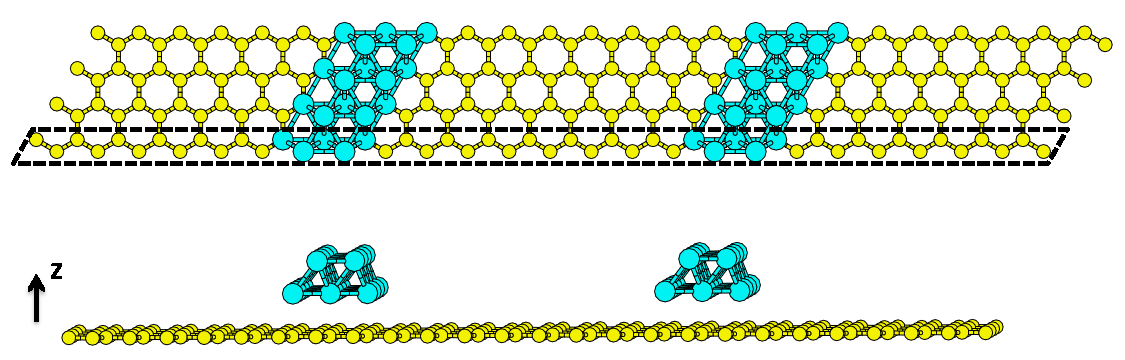}
 \end{center}
 \caption{(color online). Top and side views of the supercell (repeated by 4 times in the $\hat{y}$ direction for illustration purposes) used to calculate the magnetic coupling between two parallel cobalt ribbons (larger blue balls) placed on a graphene sheet (smaller yellow balls).}
 \label{figure2}
\end{figure}

In this section we will investigate the magnetic coupling between cobalt clusters on neutral graphene sheets which are mediated mainly by their mutual influence on the graphene $\pi$-bands. First we employ SDFT to study a relatively small system with parallel quasi-1D cobalt ribbons placed on graphene (Fig.~\ref{figure2}) and separated by $\sim 1$ nm. Then we will calculate the RKKY coupling in graphene perturbatively using the the model developed above to compare with the SDFT calculation results. This comparison informs perturbative estimates of coupling which cannot be directly addressed using {\em ab initio} tools.  

\subsection{Electronic Structure}

\begin{figure}[h]
 \begin{center}
\includegraphics[width=3.4 in]{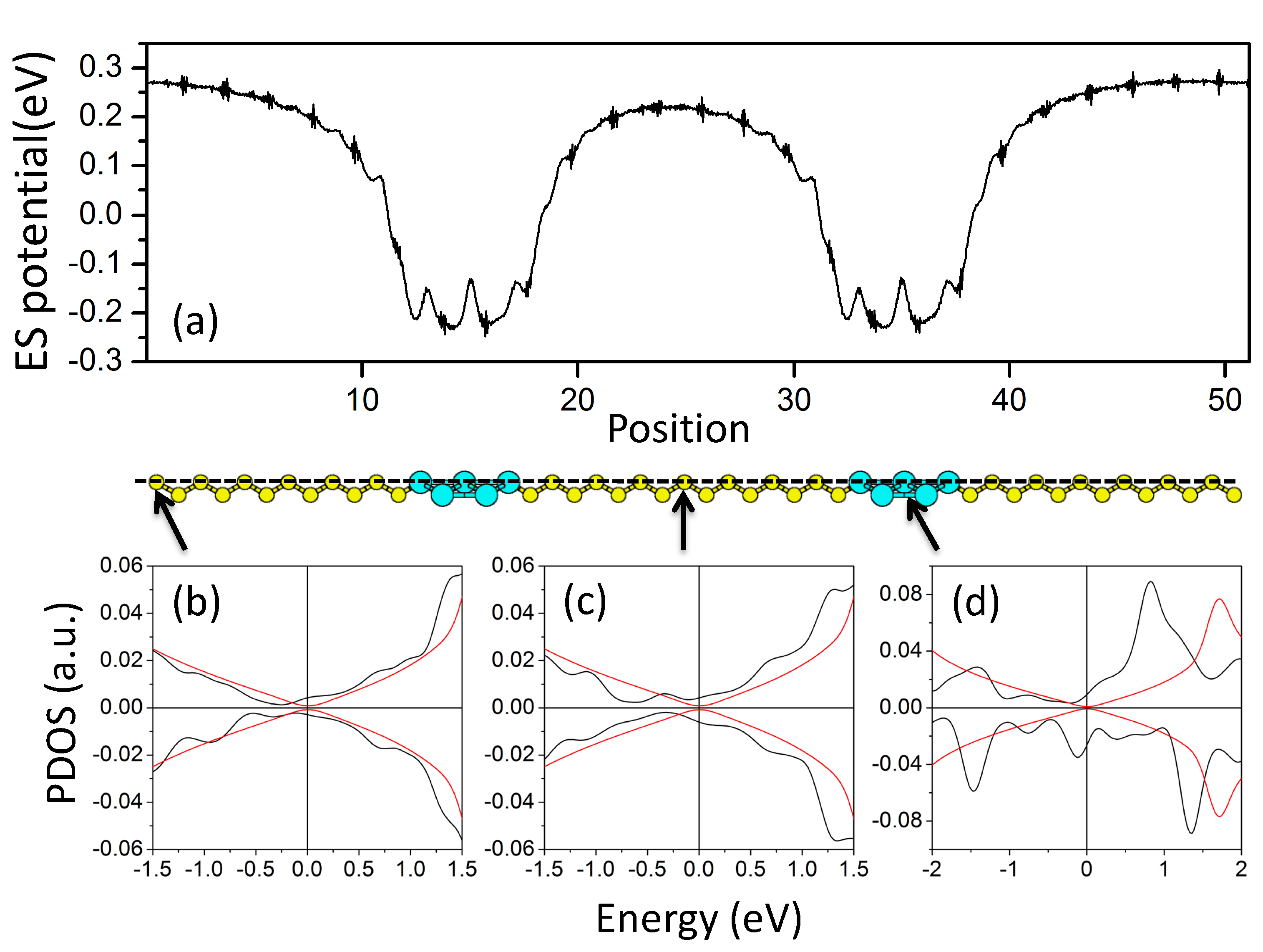}
 \end{center}
 \caption{(color online). (a) Electrostatic potential variation in adsorbed cobalt ribbons. (b-d) Density of states projected to the $p_z$ orbitals of three carbon atoms whose positions are indicated by the black arrows. Black lines--graphene with adsorbed cobalt ribbons, red lines--bare graphene. The negative PDOS axis plots minority band values while the positive axis plots majority band values.}
 \label{figure3}
\end{figure}

In Fig.~\ref{figure3} (a) we show the electrostatic potential (ionic potential plus Hartree potential from electrons) profile within the graphene sheet for the system in Fig.~\ref{figure2}. In equilibrium, the chemical potential will shift relative to the bands by the opposite amount. Therefore Fig.~\ref{figure3} (a), with a sign change and up to a constant, can be viewed as a plot of Ferm energy relative to the Dirac point. One can see that there is a large positive shift of chemical potential in the region directly below the two cobalt ribbons, meaning the graphene is strongly $n$-doped at these positions. The $\pi$-band electron barrier height between cobalt-covered and bare graphene regions is therefore about 0.5 eV, close to the 0.622 eV separation between the chemical potential and the Dirac point found earlier for the infinite 2D Co/graphene hybrid system. The barrier is smaller in the present case because separations between neighboring cobalt ribbons are not large enough for the pristine neutral graphene value. This barrier can potentially decrease magnetic coupling between remote graphene clusters by localizing electronic states more strongly in the vicinity of one particular cluster. 

In Figs.~\ref{figure3} (b-d) we plot partial density-of-states (PDOS) functions projected to the $p_z$ orbitals of carbon atoms at different points in the structure. At all three sites the PDOS Dirac-point minima are shifted to lower energy, indicating $n$-type doping over the entire graphene sheet. The magnitude of the Dirac-point shift decreases as one goes further away from the cobalt ribbons, as expected. One feature worth mentioning in the PDOS plots is the appearance of resonant features that are absent in pristine graphene. These features can be identified as confinement effects in the zigzag-ribbon-like uncovered graphene regions between the cobalt ribbons. We see later that although these modifications to the linear DOS of graphene do not greatly influence the form of the $\pi$-band mediated magnetic coupling, they do play a role in the dependence of the charge transfer to graphene on gate field.

\subsection{Exchange Coupling} 

\begin{figure}[h]
 \begin{center}
\includegraphics[width=3.4 in]{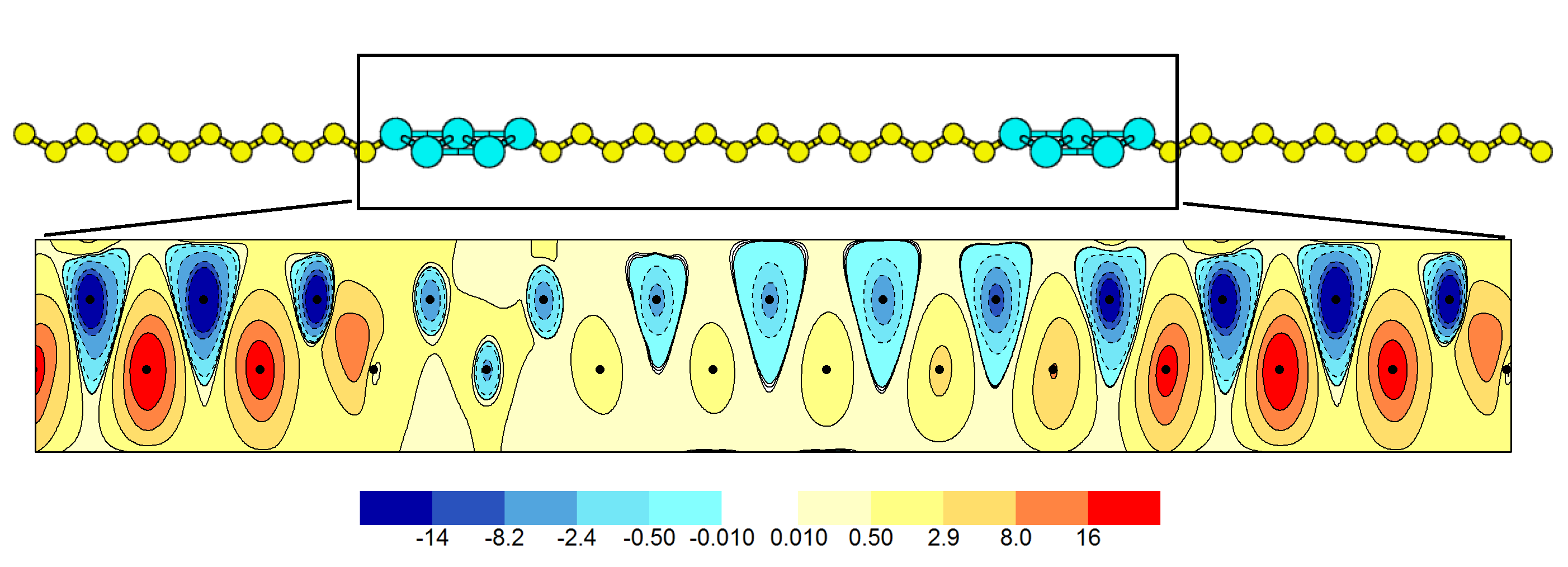}
 \end{center}
 \caption{(color online). Color scale plot of spin polarization as a function of position within the graphene plane, in the region between two cobalt ribbons with parallel spin orientations. The vertical axis in this figure is on position along the ribbon direction which has atomic scale periodicity. The positive and negative spin densities (in arbitrary units) are concentrated on carbon atoms on opposite sublattices. The black dots indicate the positions of C atoms.}
 \label{figure4}
\end{figure}

We next study the exchange coupling between the two cobalt ribbons in Fig.~\ref{figure2}. In Fig.~\ref{figure4} we plot the spin density {\em vs.} position within the graphene sheet for the case of two ferromagnetically aligned cobalt ribbons. In the region below the cobalt ribbons, the spin polarizations are opposite for the two sublattices of graphene, as in the case of complete two-layer cobalt coverage. This property is maintained in the uncovered portion of the graphene sheet. Opposite spin polarizations on the two sublattices suggests that the graphene-mediated interaction will be strongly sublattice dependent as in the RKKY case. This behavior is common in systems with bipartite lattices. \cite{saremi_2007, brey_2007}

\begin{figure}[h]
 \begin{center}
\includegraphics[width=2 in]{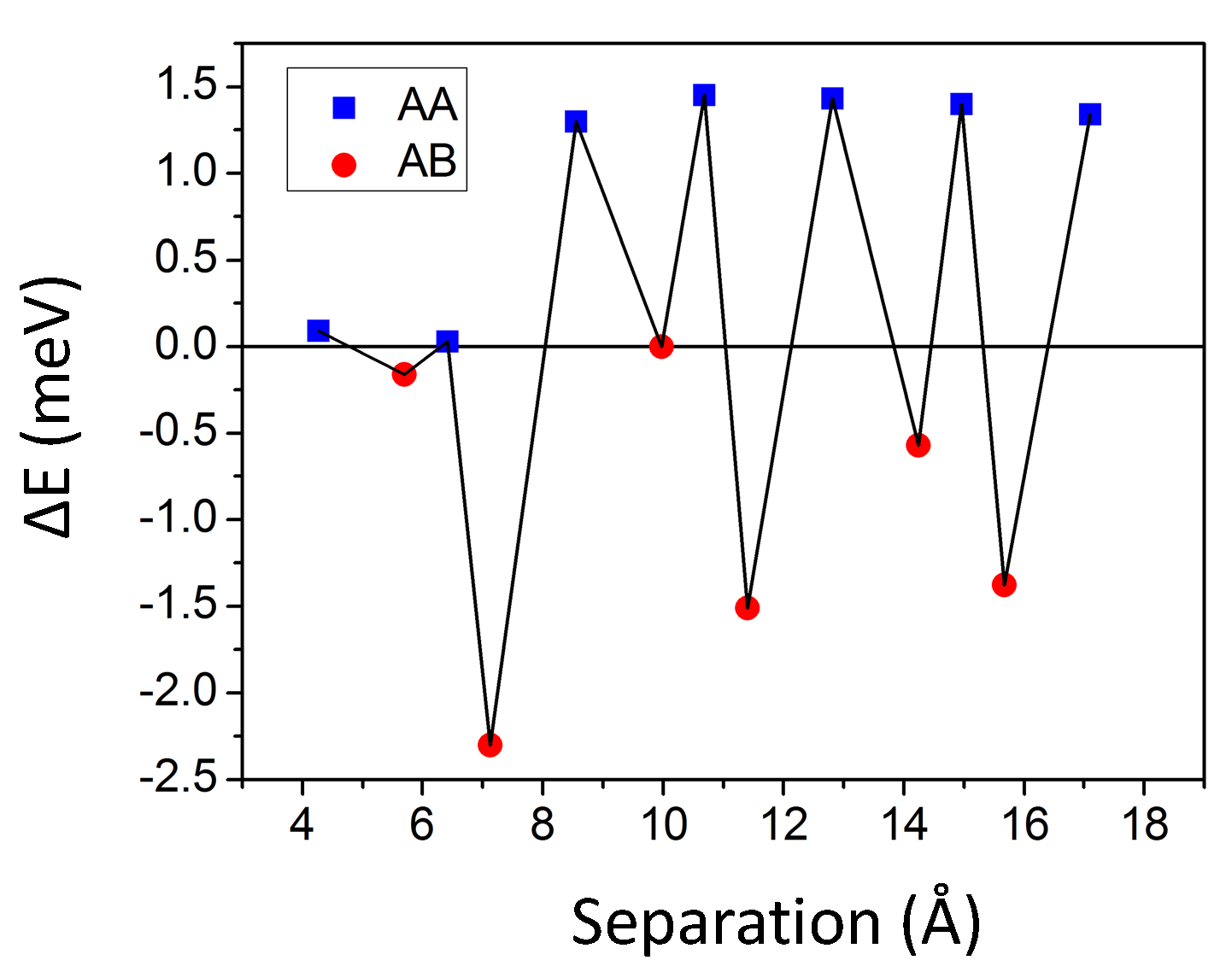}
 \end{center}
 \caption{(color online). Magnetic coupling (per supercell, which has 10 Co atoms) between two cobalt ribbons as a function of ribbon separation. The interaction strength is the total energy difference between parallel and antiparallel spin-alignment configurations. Black squares (red dots) correspond to configurations in which the cobalt atoms in the bottom layers of the two cobalt ribbons are directly above the same (different) sublattice(s) of graphene.}
 \label{figure5}
\end{figure}

In Fig.~\ref{figure5} we plot SDFT results for magnetic coupling between cobalt ribbons for different edge-to-edge separations between the ribbons and different registries with respect to the sublattices of the continuous graphene sheets. We first note that although both cobalt ribbons have the same atop-hcp registry with graphene, the first layer cobalt atom is sometimes atop an A site carbon atom and sometimes atop a B site carbon atom. The configurations of atop(A)-hcp(B) and atop(B)-hcp(A) are degenerate for an individual cobalt ribbon, but magnetic coupling energies can change if one ribbon changes registry and the other does not. The strong oscillation between FM and AFM coupling in Fig.~\ref{figure5} is due to precisely this effect. From now on we refer to the geometry in which the two cobalt ribbons have the same registry or different registries respectively as geometry AA, and geometry AB.

From Fig.~\ref{figure5} we see that the strength of the magnetic coupling is about 1.3 meV per supercell for the AA configuration for separations between 8 \r{A} and 17 \r{A}. This exchange coupling is about 0.13 meV when normalized per cobalt atom, which is much larger than the 0.04 meV MAE of bulk hcp cobalt and somewhat larger than the MAE of a 2 layer cobalt film on graphene (0.09 meV). (We have also calculated the MAE of a single cobalt ribbon on graphene as in the present setup and the value is 0.08 meV per cobalt atom, with the easy axis along the ribbon direction.) The similar strength of the MAE and the exchange coupling means that inter-ribbon interactions can have a substantial influence on the magnetic configuration of cluster arrays. RKKY-like oscillations in the coupling are expected to have period $\sim \pi/k_F$, with $k_F$ the Fermi wave vector. In the present system the Fermi energy $E_F$ is about 0.4 eV on average in the part of graphene between the two cobalt ribbons, corresponding to a period of $\sim 5$ nm. Therefore it is not surprising that we do not see RKKY-like oscillations in these calculations. The small coupling at distances below 5 \r{A} may be due to competition between direct exchange coupling and graphene-mediated coupling between the two cobalt ribbons. It is not clear why there is strong variation in the exchange coupling strength for the AB configuration. One guess is that it is due to structural details at the boundaries of the zigzag-ribbon-like graphene region between the two cobalt ribbons.

It is important for potential applications to understand how these exchange couplings will change with the size of the cobalt clusters. Due to computational power limitations we consider only two cases. First we increase the width of the two cobalt ribbons from 3 to 4 atoms, so that there are 14 cobalt atoms in a supercell. In the second case we add one more layer of cobalt atoms to the 4-atom-wide ribbons in case 1, so that the number of total cobalt atoms increases to 18. The per-cobalt magnetic coupling is 0.10 and 0.099 meV for the two cases. In both cases the per-atom coupling strength is smaller than the 0.13 meV value obtained at the original cluster size. Therefore one can expect the total exchange coupling to increase sub linearly with cluster size. There are several reasons why this finding is expected. First, as we mentioned previously, there is a large chemical potential barrier at the cluster edge, which will weaken the influence of cobalt atoms deeper inside the clusters. Second, when the cluster size is comparable to or larger than the oscillation period of the RKKY interaction, contributions from different parts of the cluster interfere destructively, as we see in the next subsection. Finally, since the largest contribution to the kinetic exchange interaction between cobalt clusters and graphene is from the cobalt atoms closest to graphene, adding more layers of cobalt to the clusters is expected to be less effective in increasing the magnetic coupling. 

\subsection{Qualitative Theory of Exchange Coupling}\label{sec:model} 

In this subsection we will use conventional perturbation theory and the model defined by Eq.~\ref{eq:model} to calculate the RKKY coupling between magnetic clusters on graphene, and compare the result with our SDFT results. Similar calculations for the RKKY interaction in graphene has been performed previously,\cite{vozmediano_2005, dugaev_2006, saremi_2007, brey_2007, bunder_2009, black-schaffer_2010, sherafati_2011_1, sherafati_2011_2, kogan_2011} but mainly for the case of point-like magnetic impurities. Here we will explicitly include the size and shape of magnetic clusters. When combined with the essential kinetic exchange parameters obtained from first principles, the formalism developed in this subsection can be a useful tool for extrapolations to system sizes beyond the range which covered by SDFT calculations.

For a graphene sheet that is partially covered by two distinct magnetic clusters 1 and 2, Eq.~\ref{eq:model} becomes  
\begin{eqnarray}
H &=& H_0+H_1+H_2\\\nonumber
 &=& \hbar v_F \hat{\bm k} \cdot \bm \tau+D_1(\bm r)\mathbb{V}_1+D_2(\bm r)\mathbb{V}_2,
\end{eqnarray}
where $D_{1(2)}(\bm r)=1$ at positions covered by cluster 1 (2) and zero otherwise,  
and $\mathbb{V}_{1(2)}=\mu_{1(2)} - h_{0,z} \tau_{z,1(2)} - h_{z,0} S_{z,1(2)} - h_{z,z} S_{z,1(2)} \tau_{z,1(2)}$. Therefore the RKKY interaction is evaluated by calculating the contribution to the total energy at second order in the perturbation $H_1+H_2$:
\begin{widetext}
\begin{eqnarray}\label{eqn:de2}
\Delta E^{(2)}=g\sum_{ss^{\prime}}\int\frac{\text{d}^2\bm k}{(2\pi)^2}\int\frac{\text{d}^2\bm k^{\prime}}{(2\pi)^2}f_{s\bm k}(1-f_{s^{\prime}\bm k^{\prime}})\frac{|\langle s\bm k|(H_1+H_2)|s^{\prime}\bm k^{\prime}\rangle |^2}{E_{s\bm k}-E_{s^{\prime}\bm k^{\prime}}} 
\end{eqnarray}
\end{widetext}
where $g=2$ is the valley degeneracy, $s=\pm 1$ is the band index,
and $f_{s\bm k}$ is the Fermi distribution function $[1+\exp((E_{s\bm k}-\mu)/k_B T)]^{-1}$. In keeping with the continuum model we are using to describe the graphene $\pi$-bands, we neglect inter-valley transitions which add an anisotropic and rapid modulation to the spatial dependence of the RKKY interaction \cite{sherafati_2011_2,kogan_2011}.

The eigenfunctions of $H_0$ are:
\begin{eqnarray}\label{eqn:Fform}
\langle\bm r|s\bm k\rangle=\frac{1}{\sqrt{2}}
\begin{pmatrix}
e^{-\text{i}\theta_{\bm k}}\\
s
\end{pmatrix}
e^{\text{i} \bm k \cdot \bm r}\equiv F_{s\bm k}e^{\text{i} \bm k \cdot \bm r} 
\end{eqnarray}
where $\theta_{\bm k}=\arctan(k_y/k_x)$. $H_{1(2)}$ can be written as a Fourier integral:
\begin{eqnarray}
H_{1(2)}(\bm r)=\int\frac{\text{d}^2 \bm q}{(2\pi)^2}e^{\text{i}\bm q \cdot \bm r}D_{{\bm q},1(2)} \mathbb{V}_{1(2)}
\end{eqnarray}
in which $D_{{\bm q},1(2)}$ is the Fourier transform of $D_{1(2)}(\bm r)$.
Therefore Eq.~\ref{eqn:de2} becomes
\begin{widetext}
\begin{eqnarray}\label{eqn:de2int}
\Delta E^{(2)} =\frac{1}{2}g\sum_{ss^{\prime}}\int\frac{\text{d}^2\bm k}{(2\pi)^2}\int\frac{\text{d}^2\bm q}{(2\pi)^2}(f_{s\bm k}-f_{s^{\prime}\bm k+\bm q})\frac{|F_{s^{\prime}\bm k+\bm q}^{\dag} (D_{{\bm q},1} \mathbb{V}_{1}+D_{{\bm q},2} \mathbb{V}_{2}) F_{s\bm k}|^2}{E_{s\bm k}-E_{s^{\prime}\bm k+\bm q}}.
\end{eqnarray}
\end{widetext}
By substituting Eq.~\ref{eqn:Fform} and the spin-dependent terms in $\mathbb{V}_{1(2)}$ into $|F_{s^{\prime}\bm k+\bm q}^{\dag} (D_{{\bm q},1} \mathbb{V}_{1}+D_{{\bm q},2} \mathbb{V}_{2}) F_{s\bm k}|^2$, and keeping only the cross terms between $D_{{\bm q},1} \mathbb{V}_{1}$ and $D_{{\bm q},2}\mathbb{V}_{2}$, we obtain 
\begin{eqnarray}
&&|F_{s^{\prime}\bm k+\bm q}^{\dag} (D_{{\bm q},1} \mathbb{V}_{1}+D_{{\bm q},2} \mathbb{V}_{2}) F_{s\bm k}|^2 = \\\nonumber
&&(D^{*}_{{\bm q},1}D_{{\bm q},2}+ {\rm c.c.}) \cdot \{ \frac{1}{2}h_{z,0}^2[1+ss^{\prime}\cos(\theta_{\bm k}-\theta_{\bm k+ \bm q})]  \\\nonumber
&&+ \frac{1}{2}h_{z,z}^2[1-ss^{\prime}\cos(\theta_{\bm k}-\theta_{\bm k+ \bm q})]  \tau_{z,1} \tau_{z,2} \}S_{z,1} S_{z,2},
\end{eqnarray} 
in which the first term in the curly brackets is sublattice-independent and the second term is sublattice-dependent. Here $\tau_{z,1(2)}$ are $\pm 1$ depending on which graphene sublattices the clusters are directly above. For conciseness we set $S_{z,1} S_{z,2} \to 1/4$ from now on. Note that the cross terms between $h_{z,0}$ and $h_{z,z}$ vanish because unperturbed graphene has spatial inversion symmetry, and $\tau_z$ changes sign under spatial inversion. Using the values of $h_{z,0}$ and $h_{z,z}$ obtained previously, the factor multiplying the sublattice-dependent term is $\sim 10$ times larger than that the factor which multiplies the sublattice-independent term. Therefore the RKKY interaction between cobalt clusters should be strongly dependent on their registration with respect to the sublattices of graphene, agreeing with our observation from the SDFT results.

The integration over $\bm k$ and the summation over bands in Eq.~\ref{eqn:de2int} can be performed explicitly at $T=0$ K. (We summarize calculation details in Appendix A.) The RKKY energy, written as an integral over $\bm q$, is 
\begin{eqnarray}\label{eqn:derkky}
\Delta E_{RKKY}^{(2)} = \frac{g h_{z,0}^2}{16\hbar v_F}\int \frac{\text{d}^2 \bm q}{(2\pi)^2} (D^{*}_{{\bm q},1}D_{{\bm q},2}+ {\rm c.c.}) \Pi_{z,0}(q) \\\nonumber
 + \frac{g h_{z,z}^2}{16\hbar v_F}\int \frac{\text{d}^2 \bm q}{(2\pi)^2} (D^{*}_{{\bm q},1}D_{{\bm q},2}+ {\rm c.c.}) \Pi_{z,z}(q) \tau_{z,1} \tau_{z,2},
\end{eqnarray}
where 
\begin{widetext}
\begin{eqnarray}
\Pi_{z,0}(q) &=& -\frac{q}{8}-\frac{k_F}{\pi}+\frac{k_F}{2\pi}\left[ \sqrt{1-\left(\frac{2k_F}{q}\right)^2}+\frac{q}{2k_F}\arcsin\frac{2k_F}{q}\right]\Theta(q-2k_F)+\frac{q}{8}\Theta(2k_F-q),\\
\Pi_{z,z}(q) &=& \frac{q}{4}-\Lambda+\frac{k_F}{\pi}-\frac{q}{2\pi}\arcsin\frac{2k_F}{q}\Theta(q-2k_F)-\frac{q}{4}\Theta(2k_F-q),
\end{eqnarray}
\end{widetext}
$\Theta(x)$ is the Heaviside step function, and $\Lambda$ is the Dirac model's ultraviolet cutoff. Note that both $\Pi_{z,0}(q)$ and its first derivative are continuous at $q=2k_F$. In contrast, $\Pi_{z,z}(q)$ has a discontinuous first derivative at $q=2k_F$, similar to the behavior of 2-dimensional electron gas. Therefore one can expect that the contribution to the RKKY interaction from the sublattice-independent part will have a faster decay with distance than that from the sublattice-dependent part.\cite{brey_2007} 

Graphene's RKKY interaction can be obtained by setting $D_1(\bm r)=\delta(\bm r)$ and $D_2(\bm r)=\delta(\bm r - \bm R)$. The $k_FR\gg 1$ limit is 
\begin{eqnarray}\label{eqn:jrkkyfinal}
J_{RKKY}(R) = \left(\frac{g h_{z,0}^2}{128 \pi \hbar v_F} - \frac{g h_{z,z}^2}{64 \pi \hbar v_F} \tau_{z,1} \tau_{z,2} \right)\cdot\frac{1}{R^3},
\end{eqnarray}
when graphene is undoped, and
\begin{eqnarray}\label{eqn:jrkkyfinal2}
J_{RKKY}(R) = - \frac{g h_{z,z}^2k_F}{16 \pi^2 \hbar v_F} \cdot \frac{\sin(2 k_F R)}{R^2} \tau_{z,1} \tau_{z,2}
\end{eqnarray}
when graphene is doped. When carriers are present the dominant contribution is the sublattice-dependent part, which is oscillatory in space and decays as $R^{-2}$. When graphene is undoped, the oscillatory term vanishes because of the $k_F$ prefactor, and the leading order terms monotonically decay as $R^{-3}$.

Next we use Eq.~\ref{eqn:derkky} to calculate the RKKY-like interaction between two cobalt ribbons on graphene with the same geometry as in our SDFT calculations. The distribution functions for this case are 
\begin{eqnarray}
D_1(\bm r) &=&  \Theta\left(x+w+\frac{d}{2}\right)\Theta\left(-x-\frac{d}{2}\right)\\\nonumber
D_2(\bm r) &=& \Theta\left(-x+w+\frac{d}{2}\right)\Theta\left(x-\frac{d}{2}\right),
\end{eqnarray}
where $d$ is the distance between the inner edges of the two ribbons, and $w$ is the width of the two ribbons. Their Fourier transforms are
\begin{eqnarray}
D_{{\bm q},1} &=& \frac{\text{i}}{q_x}\left[ e^{\text{i}q_x\frac{d}{2}}- e^{\text{i}q_x(\frac{d}{2}+w)} \right] \cdot 2\pi\delta(q_y).\\
D_{{\bm q},2} &=& \frac{\text{i}}{q_x}\left[ e^{-\text{i}q_x(\frac{d}{2}+w)} - e^{-\text{i}q_x\frac{d}{2}}\right] \cdot 2\pi\delta(q_y).
\end{eqnarray} 
Therefore,
\begin{widetext}
\begin{eqnarray}
D^{*}_{{\bm q},1}D_{{\bm q},2}+ {\rm c.c.} = \frac{2}{q_x^2}\left\{ 2\cos[q_x(d+w)]-\cos(q_x d) - \cos[q_x(d+2w)] \right\} \cdot 2\pi L \delta(q_y).
\end{eqnarray}
\end{widetext}
In deriving the above equation we have used the relation
\begin{equation}
\delta ^2(q_y)=\delta(0)\delta(q_y)=\frac{L}{2\pi}\delta(q_y),
\end{equation}
where $L$ is the length of the system in $y$ direction. We can then carry out the integration in Eq.~\ref{eqn:derkky} numerically.  Below we will compare the results from this model calculations to the SDFT results. Note that the interaction energy in SDFT is the difference between spin-parallel and spin-antiparallel configurations of the two cobalt ribbons. Therefore the model results below are all double $E_{RKKY}$ in Eq.~\ref{eqn:derkky}.

\begin{figure}[h]
 \begin{center}
\includegraphics[width=2 in]{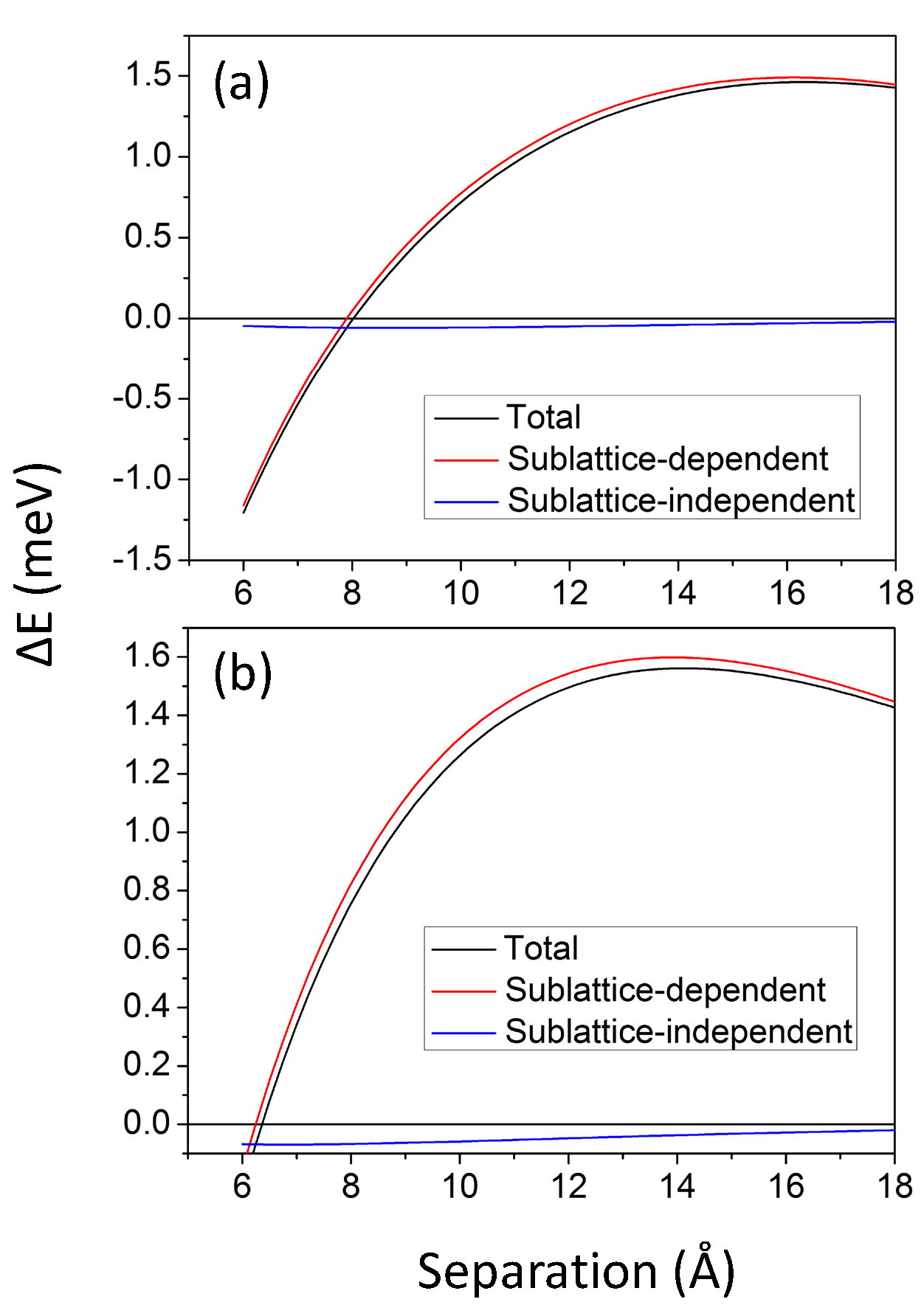}
 \end{center}
 \caption{(color online) (a) Model results for RKKY-like coupling between cobalt ribbons. $E_F$=0.4 eV, $\hbar v_F$=5.96 eV$\cdot$\r{A}, $L$=2.46 \r{A}, $w$=8.51 \r{A}. (b) Same as (a) but with $E_F$ increasing linearly as $d$ decreases.}
\label{figure6}
\end{figure}

Fig.~\ref{figure6} shows the magnetic coupling from our model for the AA geometry, which correspond to $\tau_{z,1}\tau_{z,2}=1$ in Eq.~\ref{eqn:derkky}. One can see that the order of magnitude agrees very well with the SDFT results in Fig.~\ref{figure5}, and the trend with changing distance is also well reproduced. We have chosen $E_F$ to be 0.4 eV, which is the average of the graphene chemical potential under the cobalt ribbons ($\sim$0.6 eV) and that in the center between the two cobalt ribbons ($\sim$0.2 eV). The agreement would be improved if we sued the fact that the doping level of the graphene region between the two cobalt ribbons increases as the two ribbons approach to each other. In Fig.~\ref{figure6}(b) we assumed simple linear dependence of $E_F$ with $d$ and the agreement with Fig.~\ref{figure5} is remarkably improved. We note here that there is some arbitrariness in determining the width of the cobalt ribbons $w$ since there is no sharp boundary of the portion of the graphene region which interacts with the cobalt ribbon. Here we chose $w$ to be 4 unit cells of graphene to account for the residue influence at the edges of the cobalt ribbons, although in a pure geometrical sense the cobalt ribbon amounts to 3 unit cells of graphene. $w$ may be treated as a fitting parameter in applications of our approximate theory. The magnetic coupling for the AB geometry ($\tau_{z,1}\tau_{z,2}=-1$ in Eq.~\ref{eqn:derkky}), which we did not show in Fig.~\ref{figure6}, can be obtained simply by subtracting the sublattice-dependent part from the sublattice-independent part. As we mentioned before, the anomalous oscillation of magnetic coupling for the AB geometry in Fig.~\ref{figure5} probably has a structural origin that is not captured by this simple model.

\begin{figure}[h]
 \begin{center}
\includegraphics[width=2 in]{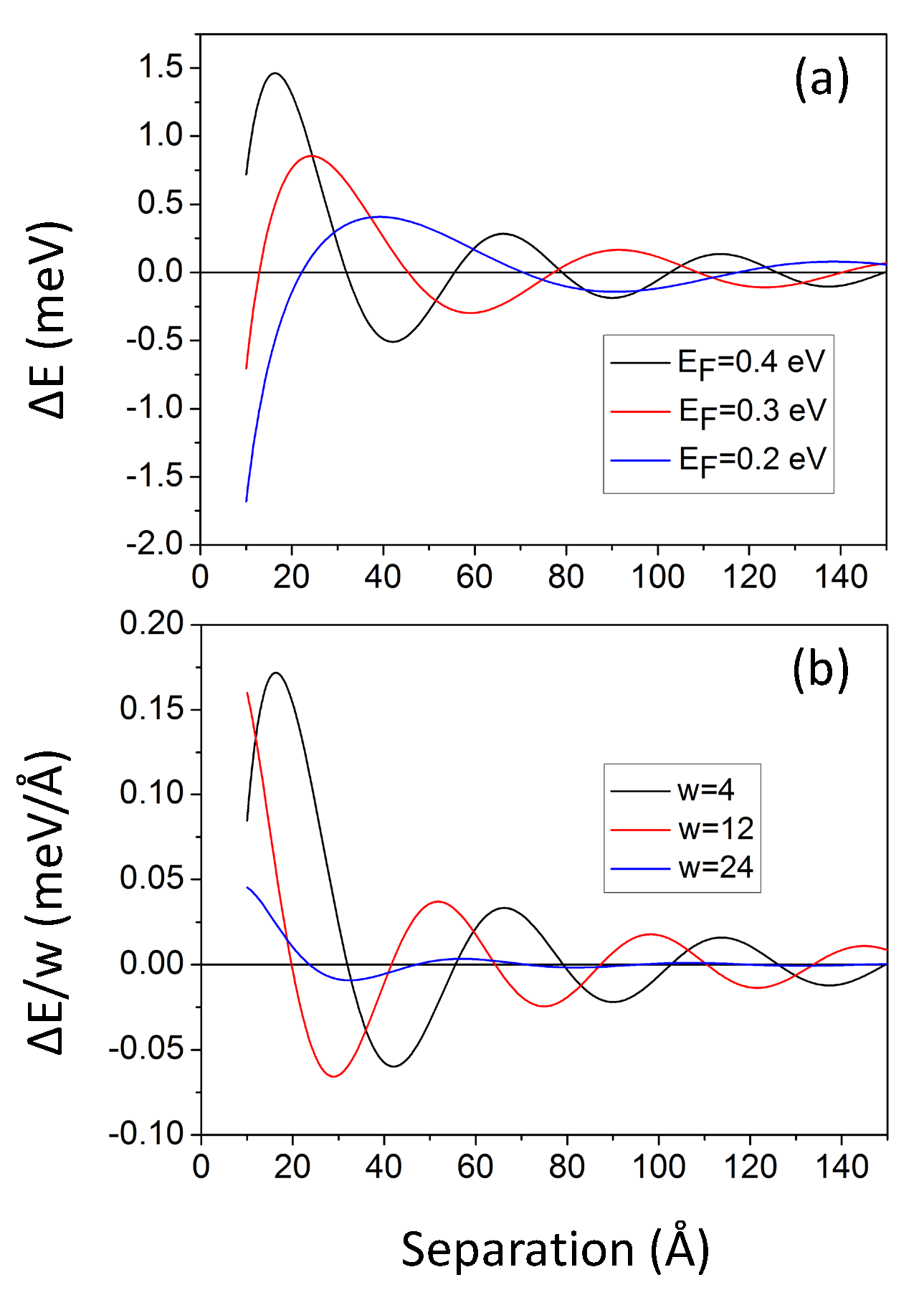}
 \end{center}
 \caption{(color online) (a) RKKY coupling between cobalt ribbons at large separations, for several carrier densities. (b) RKKY coupling divided by ribbon width $w$ for several widths. $w$ is expressed in terms of the number of graphene unit cells along the zigzag direction across the cobalt ribbon. $w$ is fixed at 4 in (a) and $E_F$ is fixed at 0.4 eV in (b).}
 \label{figure7}
\end{figure}

Knowing that our model can capture the essential physics of the graphene-mediated magnetic coupling between cobalt clusters relatively well, we can now explore the large separation limit which cannot be easily addressed by first-principles methods. First in Fig.~\ref{figure7} (a) we plot the magnetic coupling for the AA geometry {\em vs.} ribbon separation for several carrier densities. One can now see the spatial oscillation between AFM and FM interactions which appears only beyond the separation range covered in Fig.~\ref{figure5}. From the figure we see that not only the periodicity, but also the amplitude of the oscillation, depends on the doping level. This behavior is consistent with the asymptotic RKKY interaction Eq.~\ref{eqn:jrkkyfinal2}. 

In Fig.~\ref{figure7} (b) we plot magnetic coupling divided by ribbon width $w$, which is proportional to the magnetic coupling per cobalt atom. It is interesting to see that when $w$ is very large (24 graphene unit cells in the zigzag direction, equivalent to about 50 \r{A}), the magnetic coupling is strongly suppressed. This behavior can be understood by considering in terms of destructive superposition between different parts of the ribbon, when the scale of the clusters is close to the oscillation period. In addition, since the period of the RKKY oscillation increases with decreasing $k_F$, the coupling for the same large clusters will be less suppressed as the graphene is less doped, which we have also verified. Fig.~\ref{figure7} (b) also confirms our discussion on the effectiveness of increasing the magnetic coupling by preparing larger clusters. Therefore a general criterion for real applications is that the linear size of the clusters should be around or below $\frac{\pi}{2k_F}$, which is half of the RKKY period.   

\section{Gate Control of Exchange Coupling}

Since the RKKY coupling in graphene has a strong dependence on the Fermi energy (Eq.~\ref{eqn:jrkkyfinal2} and Fig.~\ref{figure7}), which in turn can be altered by electric gates, we expect that the magnetic coupling between cobalt clusters can be conveniently tuned by gating. In this section we will study the change of the magnetic coupling between cobalt ribbons on graphene with external electric fields. We have relegated some general remarks on how to simulate electric gates in supercell calculations to Appendix B.

\subsection{Freestanding Co-graphene in an Electric Field}

\begin{figure}[h]
 \begin{center}
\includegraphics[width=3 in]{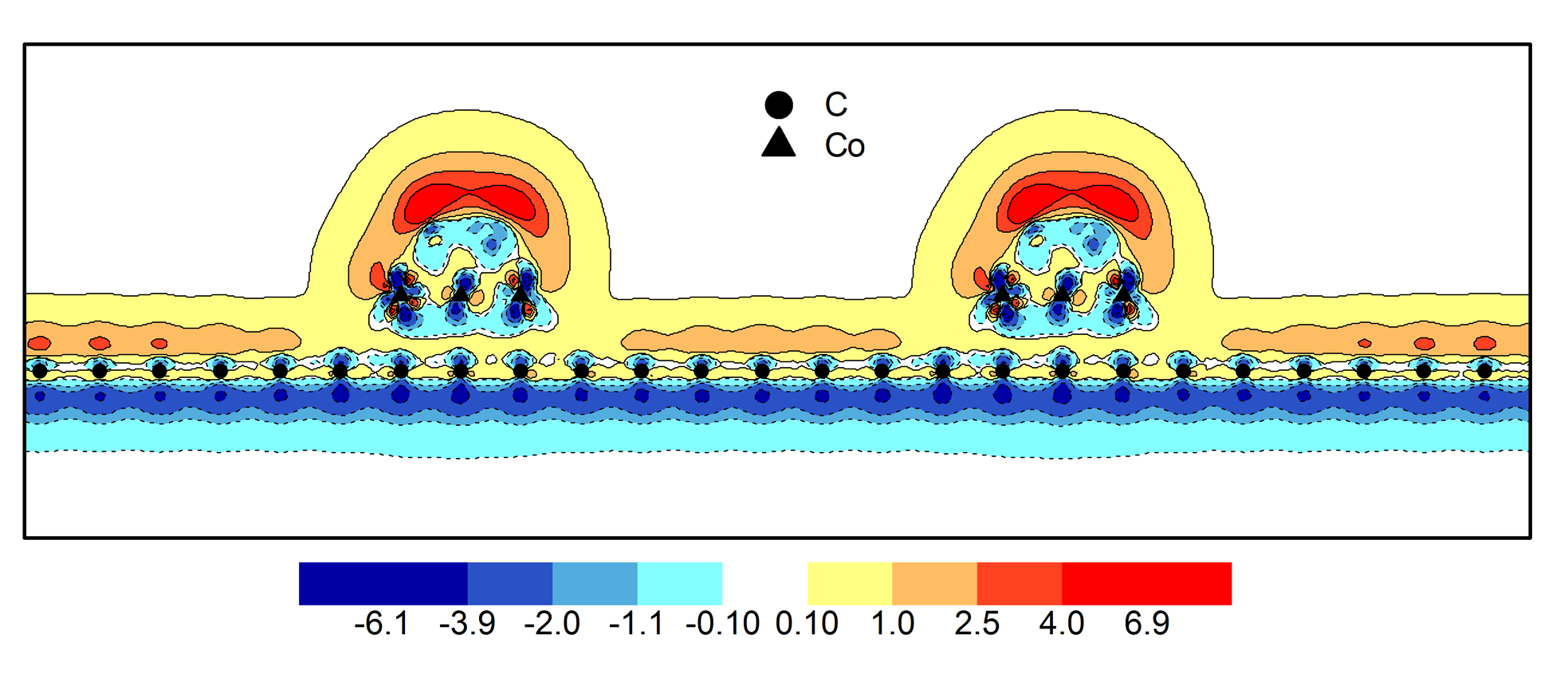}
 \end{center}
 \caption{(color online). Charge density difference (in an $x-z$ plane) between a system subjected to a 0.2 V/\r{A} electric field along the $-\hat{z}$ direction, and a system with no electric field. Positive and negative values (in arbitrary units) correspond to accumulation and depletion of charge, respectively. The black dots (triangles) indicate the positions of C (Co) atoms in the plane.}
 \label{figure8}
\end{figure}

By directly applying a electric field along the $\hat{z}$ direction in the supercell of Fig.~\ref{figure2}, we can change the Fermi energy in the graphene by transferring electrons from the cobalt ribbons to graphene and {\em vice versa}. In Fig.~\ref{figure8} we show the charge transfer within the supercell after applying a 0.2 V/\r{A} electric field along the $-\hat{z}$ direction. It can be seen that electrons are transferred from graphene to Co, and that an out-of-plane polarization is induced in the graphene sheet itself. The amount of charge transferred from the graphene plane decreases as one moves away from the cobalt ribbons, in agreement with the electrostatic potential profile shown in Fig.~\ref{figure3} (a). In this way one decreases the graphene carrier density not only in the bare regions of graphene, but also in the regions covered by the cobalt ribbons.

One question which may be raised at this point is whether or not the exchange coupling between cobalt and graphene will be influenced by the electric field. To this end we have calculated the spin polarization in a graphene sheet fully covered by a 2-layer cobalt film [Fig~\ref{figure1} (a)] under electric fields up to 0.8 V/\r{A} and did not find a significant change. Therefore the field dependence of the exchange coupling between graphene and cobalt is not an issue in the range of electric fields considered here.

\begin{figure}[h]
 \begin{center}
\includegraphics[width=3.4 in]{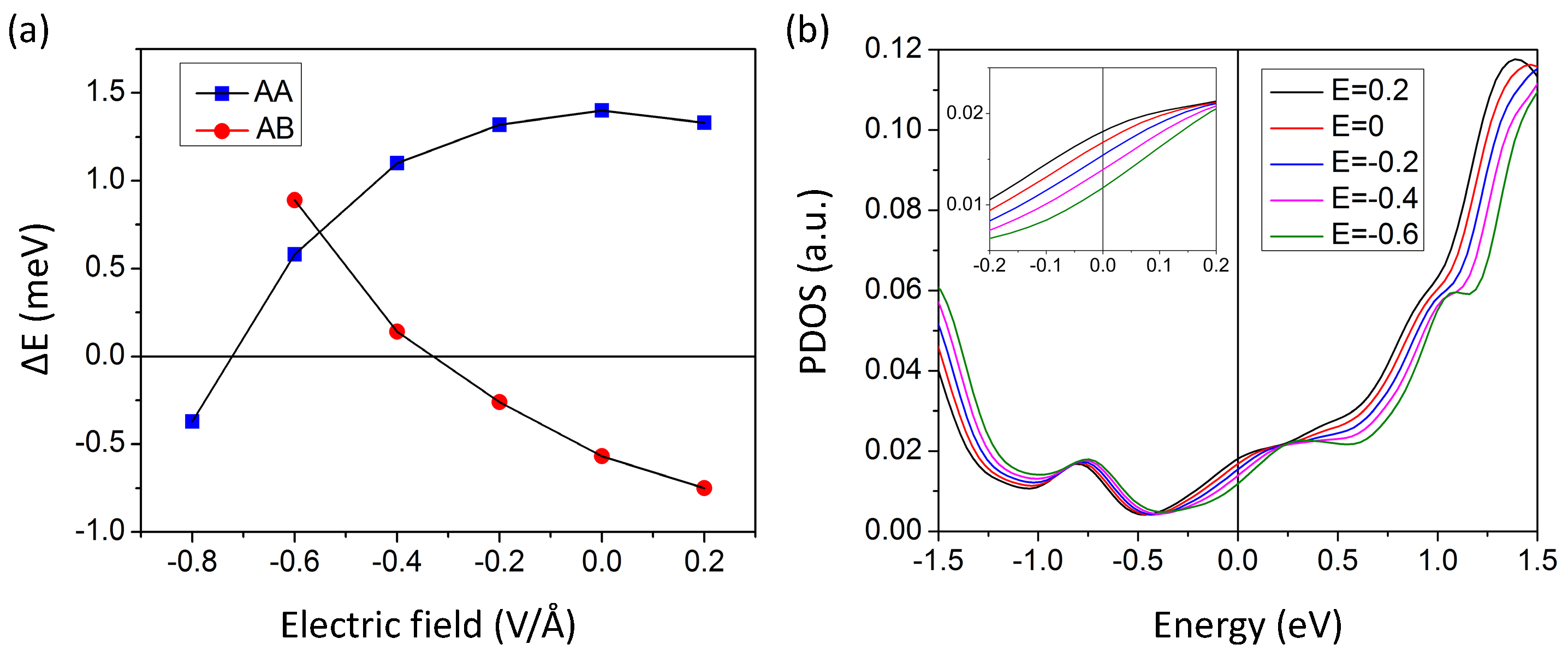}
 \end{center}
 \caption{(color online). (a) Dependence of magnetic coupling between two cobalt ribbons on external electric field at two different separations. Blue squares (red dots) correspond to the configuration that the two cobalt wires sit above the same (different) graphene sublattice(s), with a separation of 15.0 \r{A} (14.3 \r{A}). A negative value of field strength means that the field is along
the $-\hat{z}$ direction. (b) Density of states (spin-up plus spin-down) projected to the $p_z$ orbital of a C atom in the center of the supercell for several different external electric field strengths and the AA configuration in (a). The inset blows up the details around $E_F$.}
 \label{figure9}
\end{figure}

Next we study the field dependence of the magnetic coupling between the two cobalt ribbons at specific separations between them. In Fig.~\ref{figure9} (a) we plot magnetic coupling {\em vs.} electric field for two cobalt ribbons separated by $\sim$15 \r{A}, and different registries with the graphene sublattices. One can see that both the sign and magnitude of the magnetic coupling can be tuned by electric fields. It is also interesting to notice that for both the AA and AB configurations the coupling has a similar sublinear dependence on electric field. Using the simple model explained in the previous section, we found that the coupling changes almost linearly with $E_F$ from $E_F=0.2$ eV to 0.4 eV, which is roughly the range of $E_F$ shift produced by the electric fields in our DFT calculations [Fig.~\ref{figure9} (b)]. Therefore the nonlinearity should come from the field dependence of the Fermi energy of graphene. In equilibrium the external potential difference between cobalt and graphene ($e E d $ where $d$ is the spatial separation) should be balanced by the electric potential due to charge redistribution and the Fermi energy shift of graphene ({\em i.e.}, the quantum capacitance of graphene). This screening physics can be described crudely using a simple parallel plate capacitor model:
\begin{eqnarray}
e d\cdot\text{d}E=\frac{ e d c E _F \cdot \text{d}E_F}{C}+\text{d}E_F
\end{eqnarray}
where $c$ is the proportionality constant for the linear dependence of graphene DOS on $E_F$, and $c=\frac{g_vg_s}{2\pi(\hbar v_F)^2}$=0.018 eV$^{-2}$\r{A}$^{-2}$ in pure graphene, $C/d$ is the geometric capacitance of the graphene/cobalt bilayer, and $\text{d}E$ and $\text{d}E_F$ are electric field and Fermi energy differentials. The solution of this differential equation is
\begin{eqnarray}\label{eqn:EfWithE}
E_F=\frac{\sqrt{2 e^2 c d^2 C \cdot E + C^2 + 2 e c d \cdot \text{const}}-C}{e c d},
\end{eqnarray}
which explains the slower-than-linear dependence of $E_F$ on E. Of course this argument relies on the assumption that the density of states of graphene around $E_F$ is linear in energy. By looking at Fig.~\ref{figure9} (b) one can see that this assumption is actually reasonable, although the effective value of $c$ may be different from that in pure graphene value due to the confinement-induced resonances.

\begin{figure}[h]
 \begin{center}
\includegraphics[width=2 in]{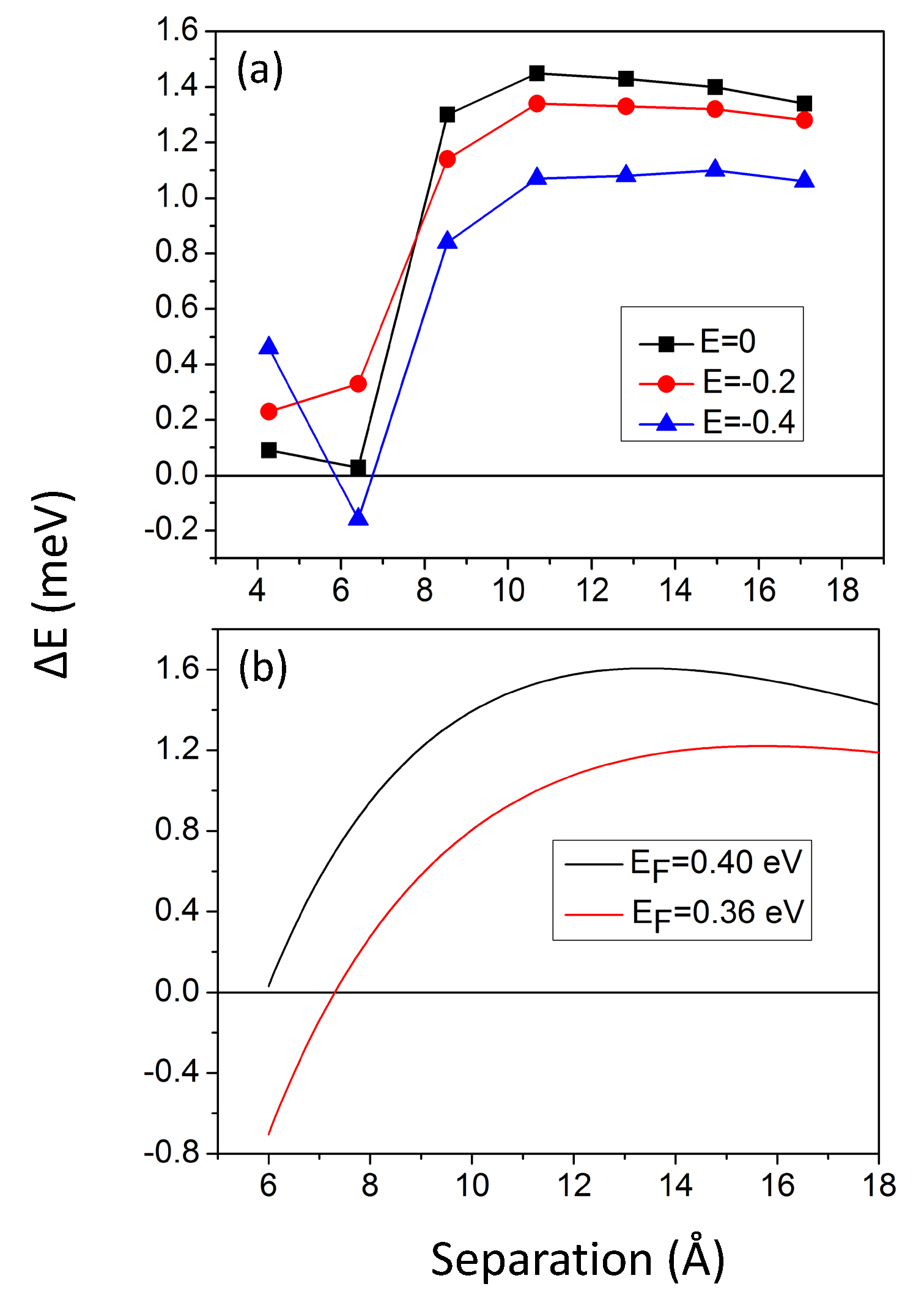}
 \end{center}
 \caption{(color online). (a) Magnetic coupling between two cobalt ribbons in the AA configuration {\em vs.} separation, under different electric fields. (b) Results obtained using the model in Sec.~\ref{sec:model}.}
 \label{figure10}
\end{figure}

Finally in Fig.~\ref{figure10} (a) we plot magnetic coupling {\em vs.} the separation between the two cobalt ribbons for several electric field strengths. The corresponding result from the model in Sec.~\ref{sec:model} is plotted in Fig.~\ref{figure10} (b). Reasonable agreement for the $E=-0.4$ V/\r{A} case is obtained by taking $E_F=0.36$ eV, which means this extremely large electric field is only able to shift $E_F$ by 0.04 eV on average. The small number is partly due to the incomplete coverage of the cobalt ribbons on graphene, which decreases the effective capacitance, but mostly due to the small vertical separation between the two systems, which makes graphene's quantum capacitance effect dominant. It is clear that an external electric field does not adequately model the influence of a remote gate. In the next subsection we will use an alternative supercell to better simulate a realistic gating geometry, and find that this tactic brings additional benefits.

\subsection{Co/graphene with a Cu Slab Mimicking a Gate Electrode}\label{sec:cuslab}

\begin{figure}[h]
 \begin{center}
\includegraphics[width=3.4 in]{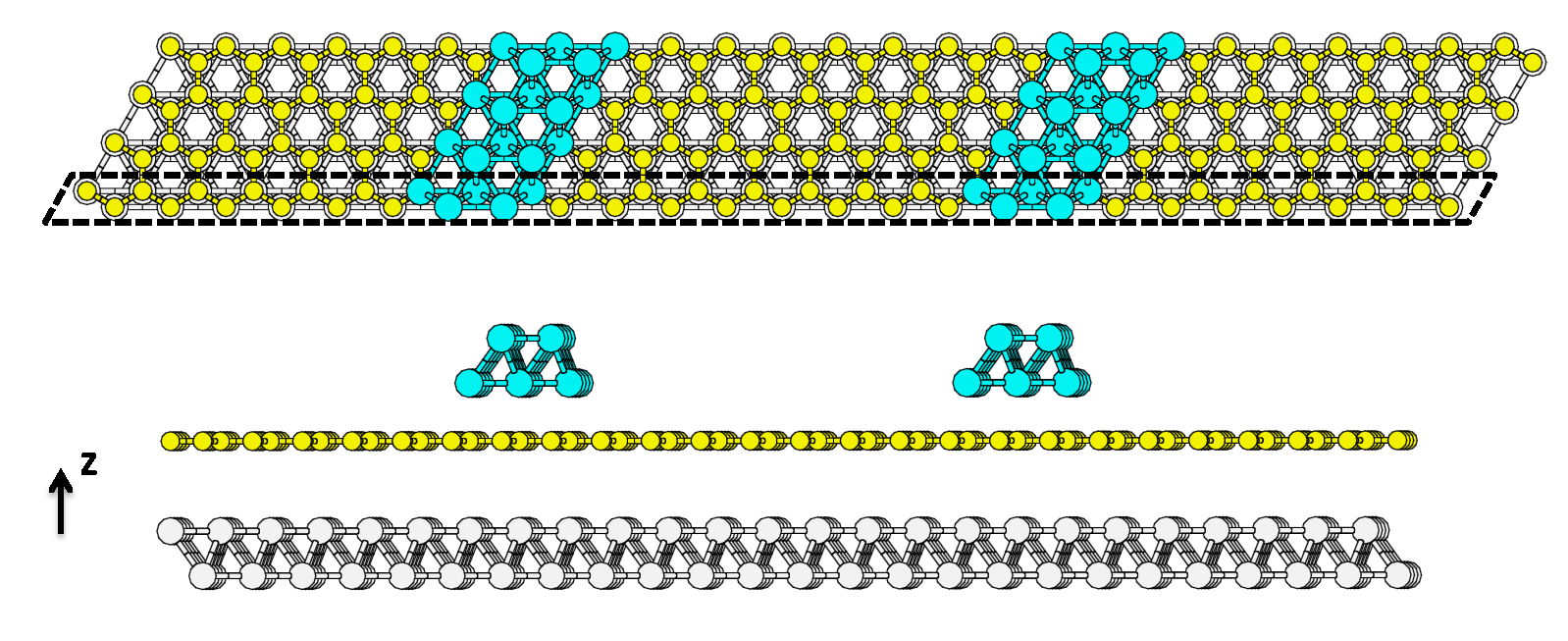}
 \end{center}
 \caption{(color online). Top and side views of the supercell with a bilayer Cu slab (grey balls) mimicking a backgate. The supercell is repeated four times in the $\hat{y}$ direction for visualization purposes.}
 \label{figure11}
\end{figure}

Fig.~\ref{figure11} shows an alternative supercell which simulates electric gating more realistically. A two atomic layer thick slab of Cu is inserted in the supercell, at a distance of about 4 \r{A} from the graphene sheet. Because of its high density of states, the Cu slab will act as an electron reservoir, just like a real gate electrode. We apply the electric field on the cobalt side of the graphene sheet and place the Cu slab on the other side of the sheet. Electrons are then transferred to or from the bare regions of graphene from the Cu slab, depending on the sign of the electric field. The part of graphene sheet that is directly below the cobalt ribbons is shielded from the the electric field by cobalt-layer screening. Consequently, complications due to field-dependent graphene cobalt coupling are mitigated. Our calculations were motivated by the expectation that adding carriers to the uncovered portion of the graphene sheet would reduce the potential barrier at the cobalt ribbon edges and in this way enhance magnetic coupling. 

\begin{figure}[h]
 \begin{center}
\includegraphics[width=3.4 in]{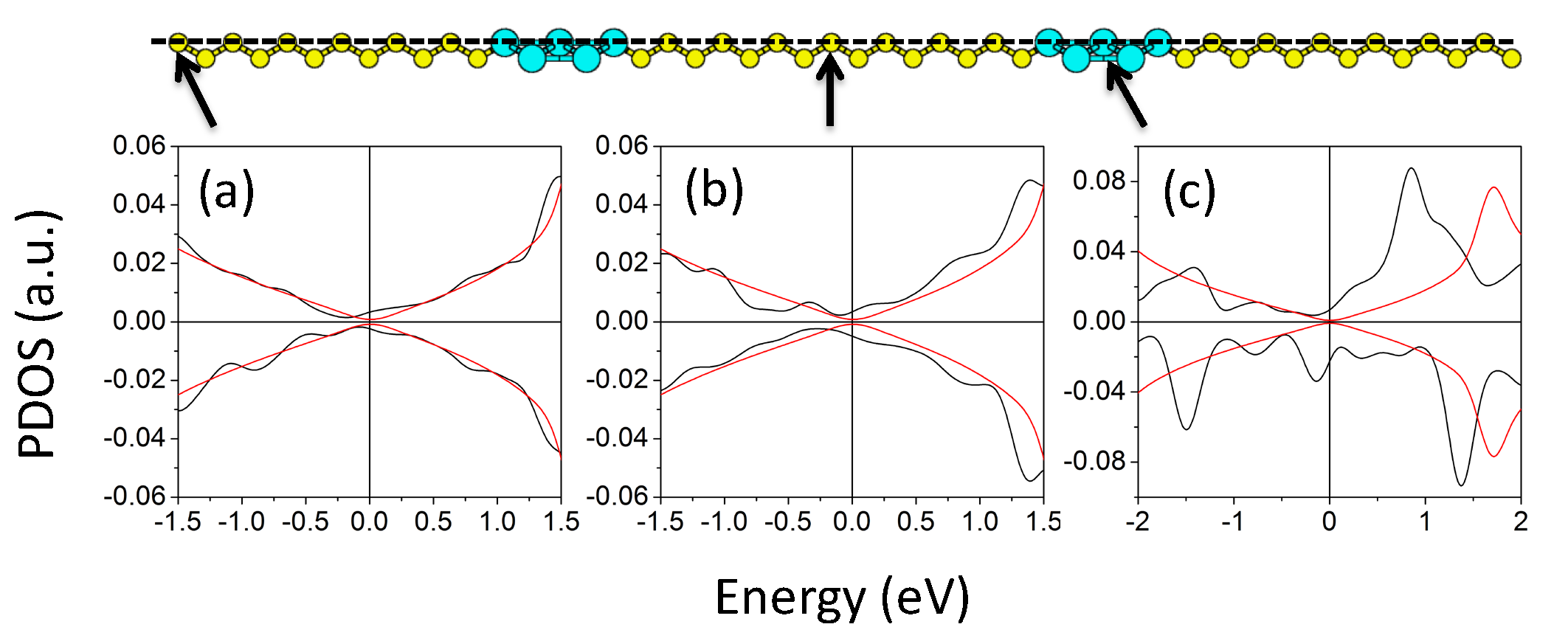}
 \end{center}
 \caption{(color online). (a-c) Density of states projected to the $p_z$ orbitals of three carbon atoms, for the supercell with a Cu slab. Black lines--graphene with the cobalt ribbons on top and the Cu slab below, red lines--bare graphene.}
 \label{figure12}
\end{figure}

In Fig.~\ref{figure12} we show PDOS for different C atoms in the graphene sheet when no external magnetic field is applied. By comparing with Fig.~\ref{figure3} one can see that the PDOS is changed mainly by a shift of $\sim$0.1 eV towards higher energies, which means that graphene is less $n$-doped. This result may seem counterintuitive since graphene is also $n$-doped on Cu, and Cu has an even smaller work function than that of Co. However, the direction of charge transfer when separation exceeds the range of direct chemical interaction is determined by relative work functions. Because Cu has a larger work function than graphene, it $p$-dopes graphene when chemically isolated \cite{khomyakov_2009}. The $p$-doping by Cu enables us to explore a doping range of graphene that cannot be easily reached by directly applying an electric field to the freestanding Co-graphene system as in the previous subsection.

\begin{figure}[h]
 \begin{center}
\includegraphics[width=3.4 in]{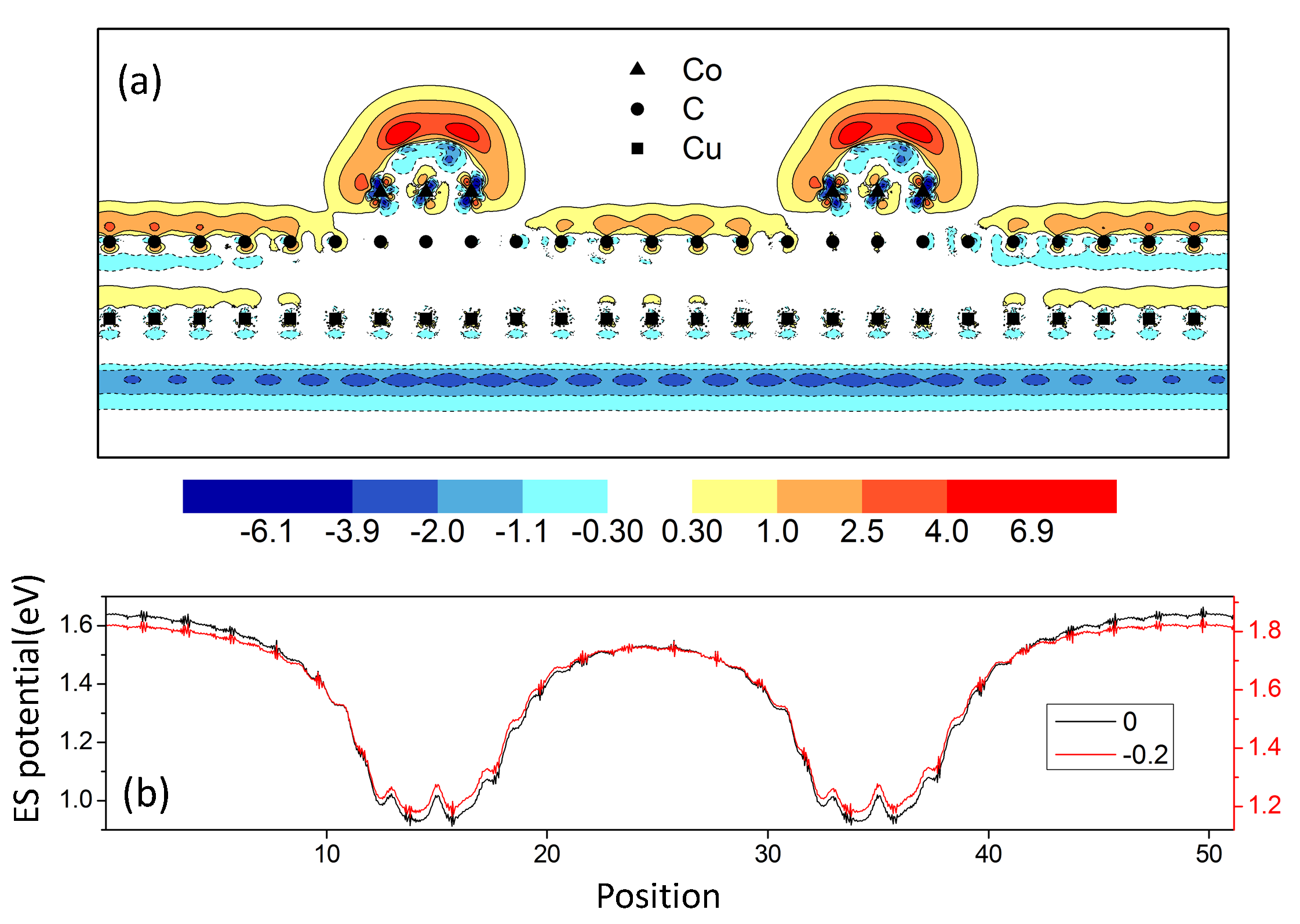}
 \end{center}
 \caption{(color online). (a) Charge density difference between the systems subjected to a 0.2 V/\r{A} electric field along the $-\hat{z}$ direction, and no electric field. Positive and negative values (in arbitrary unit) mean accumulation and depletion of charge, respectively. Black dots, triangles, and squares indicate the positions of C, Co, and Cu atoms in the plane, respectively. (b) Relative electrostatic potential as defined in Fig.~\ref{figure3} (a), for systems subjected to a -0.2 V/\r{A} electric field (red lines), and zero electric field (black lines), respectively.}
 \label{figure13}
\end{figure}

Fig.~\ref{figure13} (a) shows the charge transfer after applying a 0.2 V/\r{A} electric field along $-\hat{z}$ direction. One can see that electrons are indeed transferred from the Cu slab to the graphene and cobalt system. The part of graphene directly below the cobalt ribbons has almost no charge transfer, whereas the bare regions of graphene are electron-doped. The overall effect is essentially the same as would be produced by gating action from a planar electrode separated vertically by a distance smaller than the graphene ribbon width. From the electrostatic potential plot in Fig.~\ref{figure13} (b), the potential barriers in graphene due to the cobalt ribbons are indeed reduced after applying the field ($\sim$0.03 eV by aligning the potential at the cental region). The change is small because much of the external field is screened by the cobalt ribbons and the Cu slab. This is a limit set by our supercell size, and is therefore an artifact of our calculation procedures, but cannot be easily circumvented. Screening of the gate field due by metal clusters on graphene will however, be important experimentally when the distance to the gate is larger than the cluster separation.  

\begin{figure}[h]
 \begin{center}
\includegraphics[width=3.4 in]{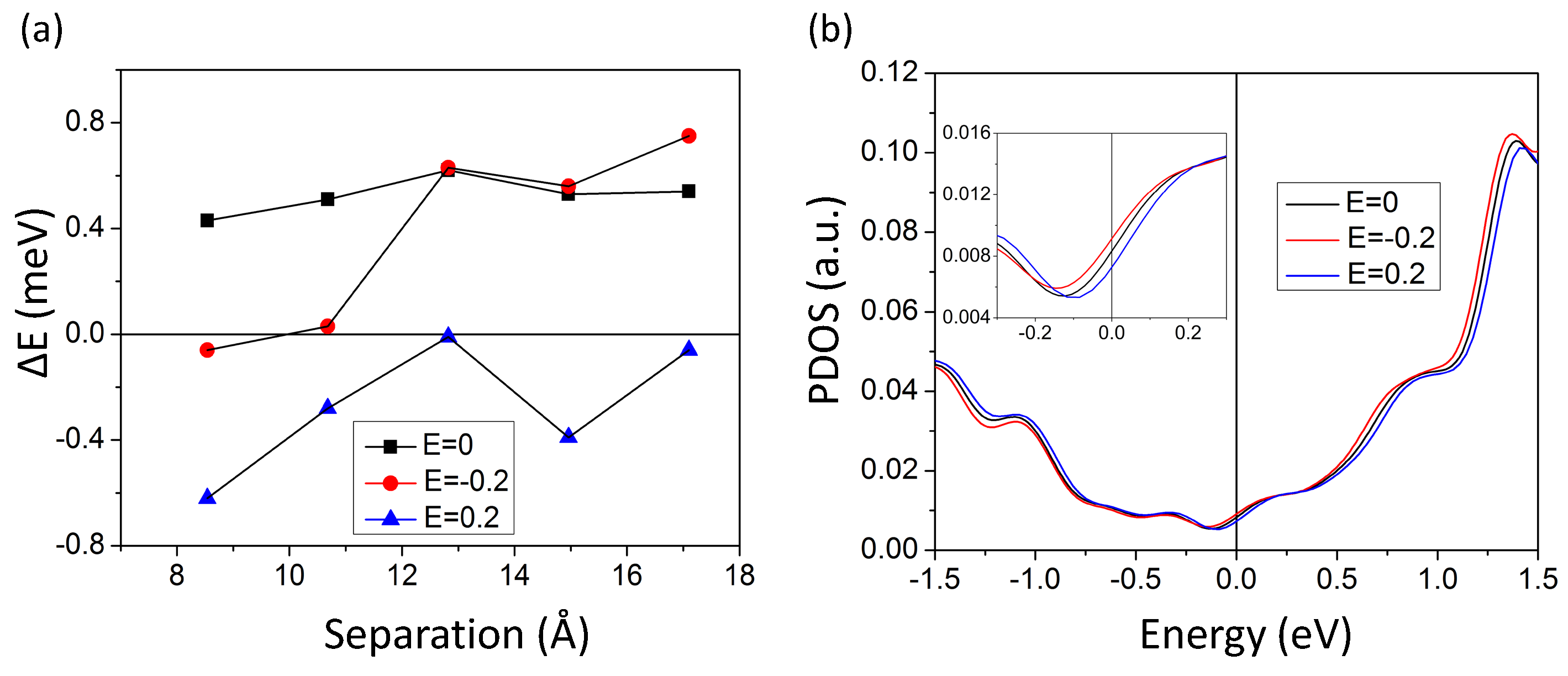}
 \end{center}
 \caption{(color online). (a) Magnetic coupling {\em vs.} separation between cobalt ribbons in the AA configuration, for several electric fields. Fields are in units of  V/\r{A}. (b) Density of states projected to the $p_z$ orbital of a C atom in the center of the supercell, with and without external electric fields, for the AA configuration and the separation of 17.1 \r{A}. The inset blows up details around $E_F$.}
 \label{figure14}
\end{figure}

Because of the different charge transfer behavior in the present supercell compared to that without the Cu slab, the field dependence of the magnetic coupling [Fig.~\ref{figure14} (a)] is changed. Without applying the electric field, the magnetic coupling is reduced because of the lower carrier density in the graphene between the two cobalt ribbons, as we have discussed previously. However, when a 0.2 V/\r{A} field is applied along the $-\hat{z}$ direction, the coupling-separation curve is changed by reduced barrier heights. Namely, when the barrier height is lower, the increase in the average doping in between the two cobalt ribbons when they get closer will be less dramatic. Since the coupling is roughly proportional to $k_F$, the shape of the coupling-distance curve should be more tilted to the left. The scenario is consistent with the model explained in Sec.~\ref{sec:model}. 

On the other hand, when a 0.2 V/\r{A} field is applied along the $z$ direction, the coupling-distance curve is relatively smooth below 13 \r{A}, a behavior which we are able to reproduce using our model. A large shift of the curve appears at around 14 \r{A}. A tentative explanation is the following: When the distance between the two cobalt ribbons is large, the central graphene region between them is nearly neutral. (In Fig.~\ref{figure14} (b) we show the PDOS of a carbon atom at the central region between the two cobalt ribbons, and it is seen that the DOS is almost linear with energy.) Therefore Eq.~\ref{eqn:EfWithE} also applies, according to which the change of $E_F$ with field will be more pronounced when $E_F$ is small. This effect, together with the fact that graphene will be more exposed to the external field as the two cobalt ribbons move away from each other, will likely lead to a sudden change of magnetic coupling at a certain separation.

\section{Discussion and Conclusions}

In this study we have demonstrated that cobalt magnetic clusters on graphene can have relatively strong gate-voltage-dependent exchange interactions, but that these interactions are sensitive to the relative sublattice registration of cobalt clusters with respect to a monolithic graphene honeycomb. Although we have focused on cobalt clusters, the combined SDFT and phenomenological modeling approach used here can be straightforwardly applied to other systems, {\em e.g.} Ni clusters on graphene.  We have carried out some similar calculations for Ni clusters, and find 
they have weaker exchange coupling with graphene than Co clusters. 
Thus cobalt has the distinct advantages of having both large exchange coupling and a good lattice match with graphene.

In Fig.~\ref{figure3} and Fig.~\ref{figure12} we have seen that resonances in the density-of-states appear due to the quantum well and edge states of the zigzag-ribbon-like uncovered graphene segments in 
our supercell calculations.  Although these density-of-states resonances do not have overwhelming importance for exchange interactions in the parameter range we were able to explore in this work, the phenomena may be interesting in 
their own right. For example, it is known that ideal graphene zigzag ribbons have spin-polarized edge states, \cite{fujita_1996, hikihara_2003, son_2006_1, son_2006_2, dutta_2008, jung_2009_1, jung_2009_2, feldner_2010, tao_2011, karimi_2012} but that graphene ribbons with impurity-free edges are very difficult, if not entirely impossible, to fabricate experimentally.\cite{tao_2011} The study of spin-polarized graphene edge states, resulting from parallel magnetic ribbons 
deposited on graphene, may be an alternative route to realizing the potentially interesting edge physics of graphene nanoribbons.
 
Our study addressed only the case of atop-hcp registry of the cobalt clusters with respect to graphene. This is the structure assumed by large 2D cobalt films on graphene. Given the strong sublattice registration dependence of this interface structure, we anticipate similar sensitivity to other structural modifications. We conclude that for any nanoparticle assembly method, precise control of the interface structure, at least in the first atomic layers, will be a crucial issue if reproduceable exchange interactions are desired. In particular, cobalt nanoparticles prepared using wet chemistry methods \cite{du_2006,petit_2004,chinnasamy_2003,antoniak_2006} are not likely to have consistent interface structures, and are therefore likely to have highly variable interactions. Here we note that some authors have concluded theoretically \cite{eom_2009, vovan_2010} that the atop-fcc interface between graphene and Co(0001) is energetically slightly preferred to atop-hcp. The difference relative to our calculations could be due to a cobalt film thickness dependence of the preferred registry, or even due to differences in the exchange-correlation potentials used in the DFT calcualtions. Nevertheless, we found that the graphene-Co exchange coupling for the atop-fcc configuration does not differ qualitatively from the atop-hcp configuration, which is expected since the dominant contribution to the exchange coupling between the Co clusters and graphene is from the atop surface Co atoms. The structural arrangement of the first row of magnetic atoms is however crucial.  

From our calculations, we can identify several key parameters that will influence the experimental realization of interesting magneto-resistance and magneto-electric devices in graphene/magnetic-metal hybrid systems. Ideally we would like to be able to substantially alter the magnetic configuration of a cluster array by changing a gate voltage. For this to happen, the inter-cluster exchange coupling should be strongly gate-voltage-dependent and the same order of magnitude as the MAE. For clusters of fixed shape, we can expect that the per-atom MAE ($\sim 10^{-4}$ eV) should be roughly cluster-size independent. The per-atom exchange coupling depends on cluster size, inter-cluster distance, and gate voltages. We can conclude that per-atom exchange coupling will be comparable to the MAE only for relatively small cluster sizes, and for relatively small inter-cluster distances. A reasonable bound for the inter-cluster distance is the period of the RKKY oscillation $\pi/k_F$, which is on the order of a few nm for graphene with a large carrier density. The cluster size also must be smaller than this number to avoid destructive superposition of coupling from different parts of a cluster. Therefore, the system size considered in our SDFT calculations is actually close to the ideal scale for strong effects. This length scale is obviously difficult to achieve, and will lead to magnetic and magneto-electric hysteresis only below $\sim 100 K$. As we mentioned in the introduction, graphene moire patterns on metal substrates provide one attractive strategy to achieve patterning on this length scale. These systems would have the disadvantage, however, that there would be no control over the relative sub lattice registration between different cobalt clusters. Another strategy is to grow large domain graphene sheets on cobalt thin films and then etch away the metal connecting different regions. In this case it should be possible to maintain control over relative sub lattice orientation, but reaching the required length scales would be challenging.     

It is interesting to compare the related case of interactions between magnetic clusters mediated by topological insulator surface states. \cite{pesin_2011, pesin_2012} In both cases the 2D metallic states are described by a Dirac model. The main differences in the topological insulator case are that graphene's sublattice degree of freedom is absent and that spin-orbit interactions are strong.  Both differences point to potential advantages of the topological insulator structures. The strong spin-orbit interactions at the TI surface will lead to strong magnetic anisotropies both in the energies of individual magnetic clusters \cite{schmidt_2011}, and in their interactions \cite{pesin_2011}, which will assist hysteresis at smaller cluster sizes. Most importantly, the absence of a sublattice degree of freedom should make it easier to control the magnetic interactions between clusters.  

In summary, we have described a survey of graphene-mediated exchange coupling between cobalt magnetic clusters, and of its tunability via electric gates. Our analysis is based on {\em ab initio} SDFT calculations interpreted using approximate models. By fitting SDFT calculations of the electronic structure of a 2D thin film of cobalt deposited on a single layer graphene sheet to a phenomenological kinetic exchange model, we have identified the relevant kinetic exchange coupling parameters. From these parameters we were able to establish that the exchange coupling between cobalt clusters is strongly sublattice registration dependent.  We then directly calculated the magnetic coupling between two infinite long cobalt ribbons on graphene using SDFT, and found that their coupling is of the same order as the magnetic anisotropy energy of the cobalt ribbons.  As expected, the coupling is found to change dramatically as one changes the relative registries of the two cobalt ribbons with the graphene sublattices. We also identified the large potential barrier at the edge of the cobalt ribbons, which may influence the magnetic coupling in a variety of ways. To explore the behaviors of the magnetic coupling in a much larger parameter range, we constructed a phenomenological theory of the magnetic coupling using the simple Dirac Hamiltonian of graphene and the kinetic exchange parameter we had obtained from the 2D calculations. The RKKY coupling given by this theory agrees well with the DFT results for the same system. We found that the magnitude of the coupling depends on the Fermi energy of graphene, and that the coupling per cobalt atom will actually be very small when the cluster size is very large. By applying an electric field inside the supercell in our SDFT calculations, we found that the electric field can lead to a considerable change in both magnitude and sign of the magnetic coupling between cobalt ribbons. The coupling changes faster with field when graphene is less doped, which was explained as a capacitance effect. We were also able to use the phenomenological theory to capture these behaviors. To better simulate the realistic gating configuration, we put a Cu slab in the supercell mimicking a backgate, which also suppresses the potential barrier at the edge of the cobalt ribbons. We found that the change of coupling with field becomes more sensitive to the separation between the two ribbons, which is a consequence of the reduced potential barriers. 

\begin{acknowledgments}
This calculation was motivated by discussions between AHM and Richard A. Kiehl during work supported by the US Army Research Office (ARO) under award number MURI W911NF-08-1-0364. AHM was supported by this award. HC would like to thank Inti Sodemann, Fengcheng Wu, Xiang Hu, Jeil Jung, Xiang Hu, Wang-Kong Tse, and Greg Fiete for valuable discussions. HC, QN, and ZZ were supported by DOE Division of Materials Sciences and Engineering Grant No. DE-FG03-02ER45958. The calculations were mainly performed at the National Energy Research Supercomputer Center (NERSC) of the US Department of Energy. 
\end{acknowledgments}

\appendix
\section{RKKY coupling of graphene from the continuous Dirac model}
\begin{widetext}
We start from calculating the integral in Eq.~\ref{eqn:de2int}, taking the sublattice-dependent term as example:
\begin{eqnarray}\label{eqn:de2ts}
\Delta E_{\tau S}^{(2)} = \frac{h_{z,z}^2}{16\hbar v_F}\sum_{ss^{\prime}}\int\frac{\text{d}^2\bm q}{(2\pi)^2} (D^{*}_{{\bm q},1}D_{{\bm q},2}+ {\rm c.c.}) \int\frac{\text{d}^2\bm k}{(2\pi)^2}(f_{s\bm k}-f_{s^{\prime}\bm k+\bm q}) \frac{1-ss^{\prime}\cos(\theta_{\bm k}-\theta_{\bm k+ \bm q})}{s|\bm k|-s^{\prime}|\bm k+\bm q|} \tau_{z,1}\tau_{z,2}.
\end{eqnarray}

We first consider the situation of $T=0$ K and $E_F$ at the Dirac point. Define
\begin{eqnarray}
\Pi^0_{z,z}({\bm q})=\sum_{ss^{\prime}}\int\frac{\text{d}^2\bm k}{(2\pi)^2}(f^0_{s\bm k}-f^0_{s^{\prime}\bm k+\bm q}) \frac{1-ss^{\prime}\cos(\theta_{\bm k}-\theta_{\bm k+ \bm q})}{s|\bm k|-s^{\prime}|\bm k+\bm q|},
\end{eqnarray}
where $f^0_{s\bm k}=\frac{1}{2}(1-s)$. To evaluate this integral we will have to calculate $\Pi^0_{z,z}(q)$ at finite freqency $\omega$:
\begin{eqnarray}
\Pi^0_{z,z}(\omega ,q) &\equiv& \sum_{ss^{\prime}}\int\frac{\text{d}^2\bm k}{(2\pi)^2}(f^0_{s\bm k}-f^0_{s^{\prime}\bm k+\bm q}) \frac{1-ss^{\prime}\cos(\theta_{\bm k}-\theta_{\bm k+ \bm q})}{s|\bm k|-s^{\prime}|\bm k+\bm q|+\omega+\text{i}\delta}\\\nonumber
 &=& -\sum_{\alpha}\int\frac{\text{d} k \text{d} \theta}{(2\pi)^2} \alpha k \frac{1+\frac{k+q\cos\theta}{|\bm k+\bm q|}}{\omega + \alpha(k+|\bm k+\bm q|)+\text{i}\delta},
\end{eqnarray}
where $\delta$ is a small real number, $\alpha=\pm 1$, and then take the limit of $\omega\rightarrow 0$. \cite{wunsch_2006} The result is
\begin{equation}\label{eqn:pi0int}
\Pi^0_{z,z}(q)= \frac{q}{4}-\Lambda,
\end{equation} 
where $\Lambda$ is a cutoff. Similarly, for the sublattice-independent part, we got
\begin{equation}\label{eqn:pi0int2}
\Pi^0_{z,0}(q)= -\frac{q}{8},
\end{equation} 
which agrees with previous results~\cite{wunsch_2006,ando_2006,brey_2007}.

Next we consider the doped case. Still take the sublattice-dependent part as example, and let 
\begin{eqnarray}
\Delta \Pi_{z,z}(q)&=&\sum_{ss^{\prime}}\int\frac{\text{d}^2\bm k}{(2\pi)^2} (\tilde{f}_{s\bm k}-\tilde{f}_{s^{\prime}\bm k+\bm q}) \frac{1-ss^{\prime}\cos(\theta_{\bm k}-\theta_{\bm k+ \bm q})}{s|\bm k|-s^{\prime}|\bm k+\bm q|}\\\nonumber
 &=& 2\sum_{ss^{\prime}}\int\frac{\text{d}^2\bm k}{(2\pi)^2} \tilde{f}_{s\bm k} \frac{1-ss^{\prime}\cos(\theta_{\bm k}-\theta_{\bm k+ \bm q})}{s|\bm k|-s^{\prime}|\bm k+\bm q|}\\\nonumber
&=& 4\int\frac{\text{d}k\text{d}\theta}{(2\pi)^2}f^{+}_{\bm k}\frac{k\cos\theta}{q+2k\cos\theta}
\end{eqnarray}
where $\tilde{f}_{s\bm k}=f_{s\bm k}-f^0_{s\bm k}$, and $f^{+}_{\bm k} = f_1(E_{1\bm k})+f_1(E_{1\bm k}+2\mu)$. The integral can be done straightforwardly. The result is
\begin{eqnarray}\label{eqn:dpiint1}
\Delta \Pi_{z,z}(q) = \frac{k_F}{\pi}-\frac{q}{2\pi}\arcsin\frac{2k_F}{q}\Theta(q-2k_F)-\frac{q}{4}\Theta(2k_F-q).
\end{eqnarray}

For the sublattice-indepedent part, after similar calculations, we got
\begin{eqnarray}
\hspace*{-1cm}\Delta \Pi_{z,0}(q) = -\frac{k_F}{\pi}+\frac{k_F}{2\pi}\left[ \sqrt{1-\left(\frac{2k_F}{q}\right)^2}+\frac{q}{2k_F}\arcsin\frac{2k_F}{q}\right]\Theta(q-2k_F)+\frac{q}{8}\Theta(2k_F-q).\label{eqn:dpiint2}
\end{eqnarray}

Next we study the behavior of the graphene RKKY interaction between two point defects, namely, the RKKY range function. The distribution function is now $D_1(\bm r)=\delta(\bm r)$ and $D_2(\bm r)=\delta(\bm r - \bm R)$, and
\begin{equation}
D^{*}_{{\bm q},1}D_{{\bm q},2}+D^{*}_{{\bm q},2}D_{{\bm q},1}=2\cos(\bm q \cdot \bm R).
\end{equation}

First we consider the sublattice-independent part. When graphene is undoped, i.e., $k_F=0$, $\Pi_{z,0}(q)=-\frac{q}{8}$ (Eq.~\ref{eqn:pi0int2}). Therefore we have
\begin{eqnarray}\label{eqn:jrkkysiint}
J_{z,0}(R) &=& -\frac{gh_{z,0}^2}{16\hbar v_F}\int \frac{\text{d}^2 \bm q}{(2\pi)^2} \frac{q \cos(\bm q \cdot \bm R) }{4}\\\nonumber
&=& -\frac{gh_{z,0}^2}{16\hbar v_F}\int \frac{\text{d} q}{2\pi} \frac{q^2}{4}J_0(qR),
\end{eqnarray}
where $J_0$ is the 0th order Bessel function. To evaluate this integral we refer to the formula~\cite{kogan_2011}
\begin{equation}
\int^{\infty}_{0}x^{n-1}e^{-px}J_{\nu}(cx)\text{d}x = (-1)^{n-1}c^{-\nu}\frac{\partial^{n-1}}{\partial p^{n-1}}\frac{(\sqrt{p^2+c^2}-p)^{\nu}}{\sqrt{p^2+c^2}}.
\end{equation}
The result is
\begin{equation}\label{eqn:jrkkysi}
J_{z,0}(R) = \frac{gh_{z,0}^2}{128 \pi \hbar v_F}\cdot\frac{1}{R^3}.
\end{equation}
Therefore at zero doping the sublattice-indepedent part corresponds to an antiferromagnetic interaction, and goes like $R^{-3}$ at large $R$. 

The situation is a little complicated when graphene is doped. Since $\Pi_{z,0}(q)$ is not singular at $q=2k_F$, the asymptotic behavior of $J_{z,0}(R)$ at large $R$ should be largely determined by the value of $\Pi_{z,0}(q)$ at small $q$. However, $\Pi_{z,0}(q)=0$ when $q<2k_F$ (Eq.~\ref{eqn:pi0int2} and \ref{eqn:dpiint2}). Therefore we can argue that $J_{z,0}(R)$ in the doped case is a superposition of two terms with similar magnitude. However, one term (corresponding to the $k_F$=0 contribution) decays monotonically as $R^{-3}$ without oscillation, while the other term will be oscillating with the periodicity related to $k_F$ since there will be singularity at $q=2k_F$ in the higher order derivatives of $\Pi_{z,0}(q)$. Therefore the long range behavior of $J_{z,0}(R)$ should still be approximately proportional to $R^{-3}$, and modulated with some oscillation. 

Finally we turn to the sublattice-dependent part $J_{z,z}(R)$
\begin{equation}
J_{z,z}(R) = \frac{gh_{z,z}^2}{16\hbar v_F}\int \frac{\text{d} q}{2\pi} 2q \Pi_{z,z}(q) J_0(qR)\tau_{z,1}\tau_{z,2}.
\end{equation}
Note that 
\begin{eqnarray}
\Pi_{z,z}(q) &=& \frac{q}{4}-\Lambda- \frac{k_F}{\pi}-\frac{q}{2\pi}\arcsin\frac{2k_F}{q}\Theta(q-2k_F)-\frac{q}{4}\Theta(2k_F-q) \\\nonumber
&=& \frac{q}{4}\left(1-\frac{2}{\pi}\arcsin\frac{2k_F}{q}\right)\Theta(q-2k_F), 
\end{eqnarray}
where we have dropped the constant terms since their Fourier transform will just be delta functions centered at $R=0$. The result of the integral is expressed in terms of the Meijer G-function:
\begin{equation}\label{eqn:meijerg}
-\frac{gh_{z,z}^2}{16\hbar v_F} \frac{1}{\pi^{\frac{3}{2}}R^3} G_{2,4}^{3,0} \left( \left. \begin{matrix} 1,1 \\ 0,\frac{3}{2},\frac{3}{2},\frac{1}{2} \end{matrix} \; \right| \, (k_F R)^2 \right).
\end{equation}
The asymptotic behavior of Meijer G-functions at large argument can be found, e.g., in Ref.~\onlinecite{meijerG}. We finally obtain the asymptotic form of Eq.~\ref{eqn:meijerg} at $k_FR\gg 1$ 
\begin{equation}\label{eqn:meijerga}
-\frac{gh_{z,z}^2}{16\hbar v_F} \frac{1}{\pi^2 R^3} \left[ \frac{3}{4} \cos(2k_F R)+k_F R \sin(2 k_F R) \right].
\end{equation} 
The asymptotic expression of the Meijer-G function turns out to work very well (Fig.~\ref{figure15}).
\begin{figure}[h]
 \begin{center}
\includegraphics[width=2 in]{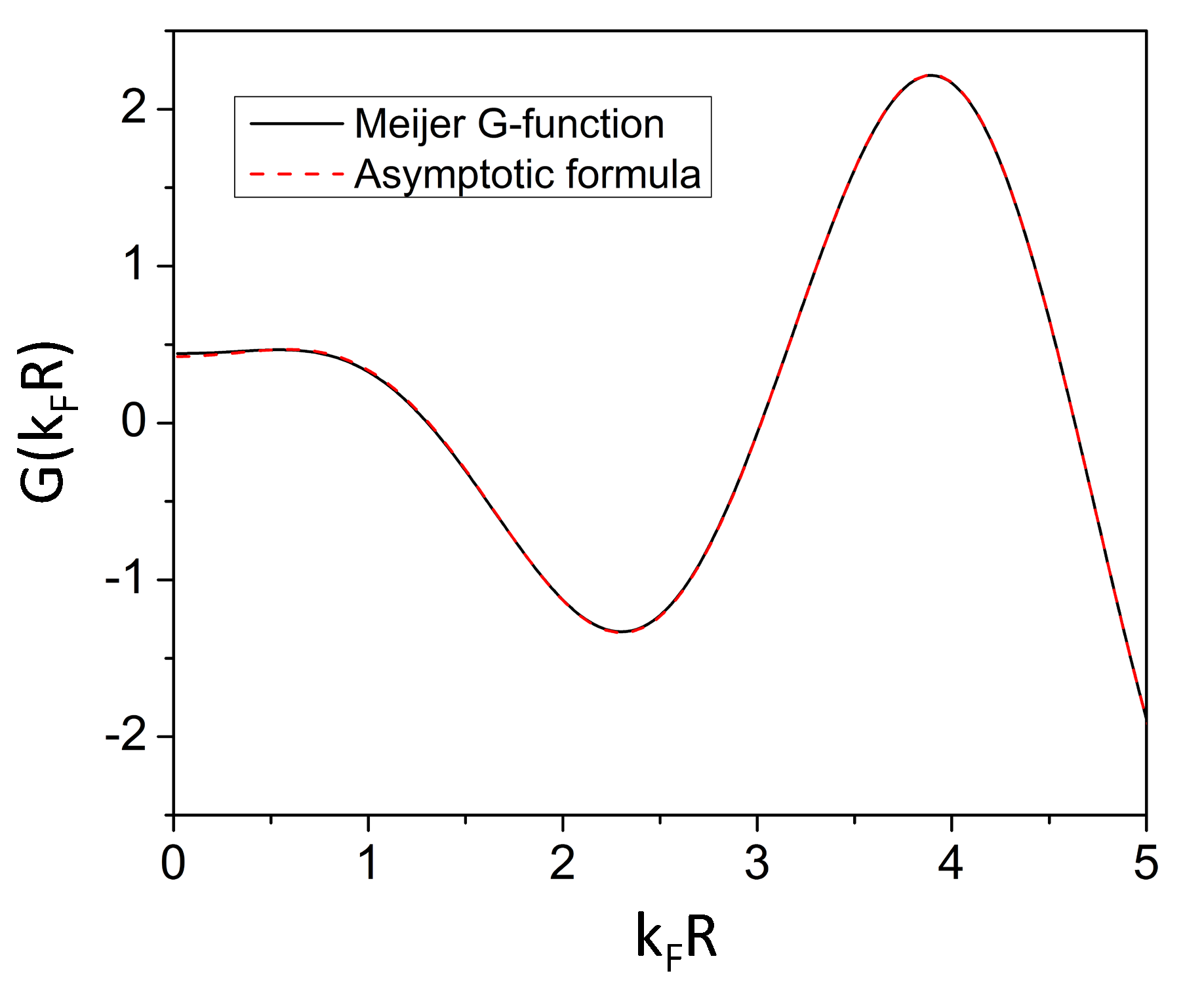}
 \end{center}
 \caption{Meijer G-function in Eq.~\ref{eqn:meijerg} and its asymptotic formula in Eq.~\ref{eqn:meijerga}.}\label{figure15}
\end{figure}
So the sublattice-dependent contribution to the RKKY interaction has an oscillating form with the period $\pi/k_F$, and the leading order term decays as $R^{-2}$, similar to the behavior of two-dimensional electron gas \cite{brey_2007,sherafati_2011_2}.  

We can finally write down the expression for the RKKY range function in graphene at $k_FR \gg 1$ by keeping only the leading order term:
\begin{eqnarray}\label{eqn:jrkkyfinal3}
J_{RKKY}(R) &=& -\frac{gh_{z,z}^2k_F}{16 \pi^2 \hbar v_F}\cdot \frac{\sin(2 k_F R)}{R^2}\tau_{z,1}\tau_{z,2}\mbox{ (doped)},\\
J_{RKKY}(R) &=&  \frac{gh_{z,0}^2}{128 \pi \hbar v_F}\cdot\frac{1}{R^3} -  \frac{gh_{z,z}^2}{64 \pi \hbar v_F}\cdot\frac{1}{R^3} \tau_{z,1}\tau_{z,2} \mbox{(undoped)}.
\end{eqnarray}
\end{widetext}

\section{Simulating gates in supercell calculations}

In this appendix we briefly discuss some of the challenges in realistically simulating gates using VASP supercell calculations. For this purpose it is natural to assume a slab geometry. \cite{gerhard_2010, negulyaev_2011}. An external potential in the supercell can be modeled straightforwardly by adding its interaction energy with electrons and ions to the Kohn-Sham energy functional. However, because the potential corresponding to a homogeneous electric field is unbounded in space, to recover the periodic boundary condition of the supercells one needs to compensate the potential difference between neighboring supercells. 
The usual scheme to do this is to add a fictitious dipole layer in the vacuum, at the boundary of the supercell \cite{neugebauer_1992}. The size of the dipole should be determined self-consistently in the minimization process of the Kohn-Sham functional, so that the dipole layer will compensate the jump of the total potential rather than that of the external potential alone. The dipole layer must be homogeneous laterally, so that it will not induce artificial fields applied to the system of interest inside the supercell. As a result, shifting the system in the supercell as a whole towards or away from the dipole layer should in principle have no impact on the properties of the system itself. In other words, the external field in the supercell is like that from two gates at plus and minus infinity, respectively. 

This feature is not desirable when one would like to simulate a circumstance in which a laterally inhomogenous system, like our graphene sheets partially covered by cobalt ribbons, that is close to a gate. The surface of a real gate is an equipotential surface, so that charge will redistribute on it when the gate is close to a system that is laterally inhomogeneous. One strategy to simulate such an equipotential boundary condition is to place a a real metal slab inside the supercell. However, attention must be paid to another difference between supercell DFT calculations and real gates, {\em i.e.} that all subsystems share the same chemical potential in the VASP case. This is a result of energy minimization in solving the Kohn-Sham equation by taking the whole supercell as one system. Consequently, spatially separate parts in the supercell act as if they were all electrically shorted. We have utilized this property in Sec.~\ref{sec:cuslab}. In the slab geometry we considered here, anything between two metal slabs (provided that they are thick enough) will be screened from external fields. Therefore, the 
best choice to simulate a real gate close to a system is to shift the system close to one boundary of the supercell, and put the metal slab at the opposite boundary from the system. We have tried this geometry using the supercells considered in this paper and found it indeed works well. The geometry, however, will not do better than the supercells used in the main text, in terms of the simulating cases with large shifts in graphene Fermi. This is because the charge redistribution on the metal slab will actually decrease the field felt by the regions of graphene not covered by cobalt ribbons.

\end{document}